\documentclass[a4paper,11pt]{article}

\usepackage{jheppub_mod} 
\usepackage[T1]{fontenc} 

\usepackage{amsmath,booktabs}
\usepackage{graphicx}
\usepackage{xspace}
\usepackage{color}
\pagecolor{white}
\usepackage[dvipsnames]{xcolor}
\usepackage{url}
\usepackage[utf8]{inputenc}
\usepackage{epstopdf}
\usepackage{hyperref}
\usepackage{multirow}

\newcommand{\beq}{\begin{equation}}
\newcommand{\eeq}{\end{equation}}
\newcommand{\beqa}{\begin{eqnarray}}
\newcommand{\eeqa}{\end{eqnarray}}

\newcommand{\mpi}{M_{\pi}}
\newcommand{\mk}{M_{K}}

\newcommand{\diff}{\text{d}}
\newcommand{\eps}{\epsilon}
\newcommand{\M}{\mathcal{M}}
\newcommand{\A}{\mathcal{A}}

\newcommand{\Order}{\mathcal{O}}

\renewcommand{\Im}{\text{Im}\,}
\renewcommand{\Re}{\text{Re}\,}

\newcommand{\Br}{\text{Br}}
\newcommand{\GeV}{\,\text{GeV}}
\newcommand{\MeV}{\,\text{MeV}}

\newcommand{\sth}{s_\text{th}}
\newcommand{\sm}{{s_\text{m}}}
\newcommand{\bsp}{\begin{sloppypar}}
\newcommand{\esp}{\end{sloppypar}}
\allowdisplaybreaks[1]

\title{Improved Standard-Model prediction for $\boldsymbol{K_L\to \ell^+\ell^-}$}

\author[a]{Martin Hoferichter,}
\author[a]{Bai-Long Hoid,}
\author[b]{and Jacobo Ruiz de Elvira}

\affiliation[a]{Albert Einstein Center for Fundamental Physics, Institute for Theoretical Physics, University of Bern, Sidlerstrasse 5, 3012 Bern, Switzerland}

\affiliation[b]{Departamento de F\'isica Te\'orica and IPARCOS, Facultad de Ciencias F\'isicas,
Universidad Complutense de Madrid, Plaza de las Ciencias 1, 28040 Madrid, Spain}

\emailAdd{hoferichter@itp.unibe.ch}
\emailAdd{longbai@itp.unibe.ch}
\emailAdd{jacobore@ucm.es}

\abstract{We present a comprehensive calculation of the $K_L\to\gamma^*\gamma^*$ form factor in dispersion theory, using input from the leptonic decays $K_L\to\ell^+\ell^-\gamma$, $K_L\to \ell_1^+\ell_1^-\ell_2^+\ell_2^-$, the hadronic mode $K_L\to \pi^+\pi^-\gamma$, the normalization $K_L\to\gamma\gamma$, and the matching to asymptotic constraints. As key result we obtain an improved determination of the long-distance contribution to $K_L\to\ell^+\ell^-$, leading to the Standard-Model predictions $\Br[K_L\to\mu^+\mu^-]=7.44^{+0.41}_{-0.34}\times 10^{-9}$, $\Br[K_L\to e^+e^-]=8.46(37)\times 10^{-12}$, and more stringent limits on physics beyond the Standard Model. We provide a detailed breakdown of the current uncertainty, and delineate how future experiments and the interplay with lattice QCD could help further improve the precision.}

\begin{document}

\preprint{IPARCOS-UCM-23-089}

\maketitle
	
\section{Introduction}
\label{sec:intro}

Rare kaon decays are sensitive low-energy probes of physics beyond the Standard Model (BSM)~\cite{Cirigliano:2011ny}. Next to the neutrino modes $K^+\to\pi^+\nu\bar\nu$~\cite{NA62:2021zjw} and $K_L\to\pi^0\nu\bar\nu$~\cite{KOTO:2018dsc}, another promising channel concerns the decay of a neutral kaon into a dilepton pair, whose short-distance (SD) properties differ markedly for the $K_S$ and $K_L$ channels. The general decomposition of the amplitude 
\beq
\label{BC}
\A\big[K\to\ell^+(p_1)\ell^-(p_2)\big]=\bar u(p_2)\big[i B+C\gamma_5\big]v(p_1)
\eeq
leads to a decay width
\beq
\Gamma\big[K\to\ell^+\ell^-\big]=\frac{\mk}{8\pi}\sigma_\ell(\mk^2)\big[(\sigma_\ell(\mk^2))^2|B|^2+|C|^2\big],\qquad \sigma_\ell(\mk^2)=\sqrt{1-\frac{4m_\ell^2}{\mk^2}}.
\eeq
For $K_S\to\ell^+\ell^-$ the dominant contribution arises from the long-distance (LD) $K_S\to\gamma^*\gamma^*\to\ell^+\ell^-$ $P$-wave amplitude $B$~\cite{DAmbrosio:1986zin,Ecker:1991ru}, which, due to the absence of $CP$-invariant local contributions at two-loop order in chiral perturbation theory (ChPT) can be predicted with reasonable accuracy~\cite{Ecker:1991ru}
\beq
\label{KS_LD}
\Br[K_S\to\mu^+\mu^-]\big|_\text{LD}\simeq 5.0(3)\times 10^{-12},\qquad 
\Br[K_S\to e^+e^-]\big|_\text{LD}\simeq 2.1(1)\times 10^{-14},
\eeq
where the uncertainty is propagated from $\Br[K_S\to\gamma\gamma]=2.63(17)\times 10^{-6}$~\cite{ParticleDataGroup:2022pth,KLOE:2007rta,Lai:2002sr}, but does not include theory uncertainties in $\Br[K_S\to\ell^+\ell^-]/\Br[K_S\to\gamma\gamma]$ itself.
Current limits~\cite{LHCb:2020ycd,KLOE:2008acb}
\beq
\Br[K_S\to\mu^+\mu^-]<2.1\times 10^{-10},\qquad 
\Br[K_S\to e^+e^-]<9\times 10^{-9},
\eeq
are less than two orders of magnitude away for the muon mode, while the electron channel appears out of reach for the foreseeable future. In this process, a $CP$-violating SD contribution occurs via the $S$-wave  amplitude $C$, at the level of $\Br[K_S\to\mu^+\mu^-]\big|_\text{SD}\simeq 1.7(2)\times 10^{-13}$~\cite{Buchalla:1995vs,Isidori:2003ts,Cirigliano:2011ny,Brod:2022khx}. A first step towards an improved calculation of $\pi\pi$ rescattering effects in the LD SM contribution was taken in Ref.~\cite{Colangelo:2016ruc} (in terms of partial waves for $\gamma^*\gamma^*\to\pi\pi$~\cite{Garcia-Martin:2010kyn,Hoferichter:2011wk,Moussallam:2013una,Danilkin:2018qfn,Hoferichter:2019nlq,Danilkin:2019opj}), which would allow one to reduce the hidden uncertainty in Eq.~\eqref{KS_LD} due to higher orders in ChPT~\cite{DAmbrosio:2022kvb}. 

\begin{figure}[t]
	\centering
	\includegraphics[width=0.35\linewidth]{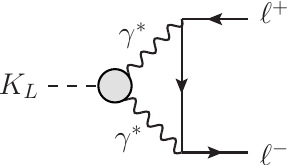}
	\caption{LD contribution to $K_L\to\ell^+\ell^-$. The gray blob refers to the $K_L\to\gamma^*\gamma^*$ transition form factor.} 
	\label{fig:diagram_LD}
\end{figure}

\begin{figure}[t]
	\centering
	\includegraphics[height=.9\linewidth,angle=90]{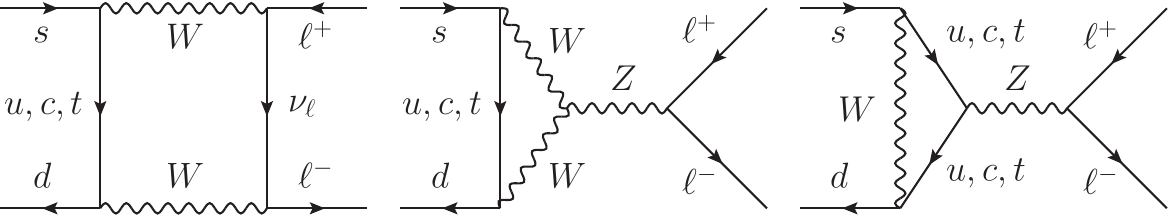}
	\caption{SD contributions to $K_L\to\ell^+\ell^-$ in the SM.} 
	\label{fig:diagrams_SD}
\end{figure}

For $K_L\to\ell^+\ell^-$ the roles become reversed, and the $CP$-conserving LD contribution now proceeds via the $C$ amplitude in Eq.~\eqref{BC}. Normalizing to the $K_L\to\gamma\gamma$ decay, the decay rate can be written as
\beq
\label{R_ell}
R_L^\ell=\frac{\Br[K_L\to\ell^+\ell^-]}{\Br[K_L\to\gamma\gamma]}=2\sigma_\ell(\mk^2)\bigg(\frac{\alpha}{\pi}r_\ell\bigg)^2\big|\A_\ell(\mk^2)\big|^2,\qquad r_\ell = \frac{m_\ell}{\mk},\qquad \alpha=\frac{e^2}{4\pi},
\eeq
where the dominant imaginary part comes from $\gamma\gamma$ intermediate states~\cite{Martin:1970ai}
\beq
\text{Im}_{\gamma\gamma}\A_\ell(\mk^2)=\frac{\pi}{2\sigma_\ell(\mk^2)}\log\big[y_\ell(\mk^2)\big],\qquad y_\ell(\mk^2)=\frac{1-\sigma_\ell(\mk^2)}{1+\sigma_\ell(\mk^2)}.
\eeq
The real part is typically decomposed into the loop function that corresponds to the one-loop result in ChPT and a local contribution $\chi(\mu)$~\cite{GomezDumm:1998gw,Knecht:1999gb,Isidori:2003ts} 
\beq
\label{ReA}
\Re\A_\ell(\mk^2)=\frac{1}{\sigma_\ell(\mk^2)}\bigg[\text{Li}_2\big[-y_\ell(\mk^2)\big]+\frac{1}{4}\log^2\big[y_\ell(\mk^2)\big]+\frac{\pi^2}{12}\bigg]+3\log\frac{m_\ell}{\mu}-\frac{5}{2}+\chi(\mu),
\eeq
where $\chi(\mu)$ receives both LD and SD contributions in the SM,\footnote{We follow the convention of Refs.~\cite{Knecht:1999gb,Isidori:2003ts} for $\chi(\mu)$, which is related to the one of Refs.~\cite{GomezDumm:1998gw,Cirigliano:2011ny} by $\chi(\mu)\big|_\text{\cite{Knecht:1999gb,Isidori:2003ts}}=\chi(\mu)\big|_\text{\cite{GomezDumm:1998gw,Cirigliano:2011ny}}+5/2$, and yet other choices exist depending on whether the tadpole subtraction or strict $\overline{\text{MS}}$ are used in the ChPT calculation~\cite{Savage:1992ac,Ametller:1993we}. The variant in Eq.~\eqref{ReA} allows for a direct comparison to recent calculations of $\pi^0\to e^+e^-$~\cite{Vasko:2011pi,Husek:2014tna,Hoferichter:2021lct}.} see Figs.~\ref{fig:diagram_LD} and~\ref{fig:diagrams_SD}.
Fixing the sign following the arguments from Refs.~\cite{Pich:1995qp,GomezDumm:1998gw,Isidori:2003ts}, the SD part is known to about $4\%$ precision~\cite{Buchalla:1993wq,Gorbahn:2006bm}
\beq
\label{chi_SD_SM}
\chi_\text{SD}^\text{SM}=-1.80(6),
\eeq
see App.~\ref{app:SD}, with uncertainties dominated by CKM matrix elements.
The experimental results~\cite{ParticleDataGroup:2022pth,E871:2000wvm,Akagi:1994bb,E791:1994xxb,BNLE871:1998bii}
\beq
\Br[K_L\to\mu^+\mu^-]=6.84(11)\times 10^{-9},\qquad \Br[K_L\to e^+ e^-]=8.7^{+5.7}_{-4.1}\times 10^{-12},
\eeq
or, using $\Br[K_L\to\gamma\gamma]=5.47(4)\times 10^{-4}$~\cite{ParticleDataGroup:2022pth},\footnote{For $R_L^\mu$ we use $\Gamma[K_L\to\mu^+\mu^-]/\Gamma[K_L\to\pi^+\pi^-]=3.477(53)\times 10^{-6}$~\cite{E871:2000wvm,Akagi:1994bb,E791:1994xxb}, together with $\Gamma[K_L\to\pi^+\pi^-]/\Gamma[K_L\to\gamma\gamma]=3.596(44)$ from the global fit of Ref.~\cite{ParticleDataGroup:2022pth} (including correlations).}
\beq
\label{RLexp}
R_L^\mu=1.250(24)\times 10^{-5},\qquad R_L^e=1.59^{+1.04}_{-0.75}\times 10^{-8},
\eeq
would thus allow one to derive BSM constraints if the LD contribution to $\chi(\mu)$ could be calculated. 

Previous theoretical estimates of this contribution have relied on ChPT, leading to diagrams including the mixing of the $K_L$ to $\pi^0$, $\eta$, $\eta'$~\cite{GomezDumm:1998gw,Knecht:1999gb,Greynat:2003ja}, or on parameterizations of the $K_L\to\gamma^*\gamma^*$ form factor informed by the radiative decays $K_L\to\ell^+\ell^-\gamma$~\cite{DAmbrosio:1997eof,Valencia:1997xe} and the asymptotic behavior expected from a partonic calculation~\cite{Isidori:2003ts}. More recently, also first steps towards a lattice-QCD calculation of $\chi(\mu)$ have been taken~\cite{Chao:2023cxp,Christ:2022rho,Zhao:2022pbs,Christ:2020bzb}, and proposals were developed to extract SD information from the time evolution of the $K\to\ell^+\ell^-$ decay rate~\cite{DAmbrosio:2017klp,Dery:2021mct,Dery:2022yqc}. In view of the experimental prospects for future precision measurements of $K_L$ decays in phase 2 of HIKE~\cite{HIKE:2022qra,HIKE:2023ext} and at KOTO II~\cite{Aoki:2021cqa,Nanjo:2023xvj} (see also Ref.~\cite{NA62KLEVER:2022nea,Anzivino:2023bhp}), it is thus timely to revisit the calculation of the $K_L\to\gamma^*\gamma^*$ form factor using methods that were recently developed in the context  of pseudoscalar-pole contributions to hadronic light-by-light scattering in the anomalous magnetic moment of the muon~\cite{Aoyama:2020ynm,Colangelo:2014dfa,Colangelo:2014pva,Colangelo:2015ama,Hoferichter:2018dmo,Hoferichter:2018kwz}. 

To this end, we analyze the $K_L\to\gamma^*\gamma^*$ form factor in dispersion theory, which allows us to not only include the leptonic modes $K_L\to\ell^+\ell^-\gamma$, $K_L\to\ell_1^+\ell_1^-\ell_2^+\ell_2^-$ in its reconstruction from data, but, crucially, also the hadronic channel $K_L\to\pi^+\pi^-\gamma$. In addition, we perform the matching to asymptotic constraints via a dispersive representation of the respective loop integrals, which allows for a better separation of their low- and high-energy components. This strategy follows closely our calculation of the $\pi^0\to\gamma^*\gamma^*$ transition form factor~\cite{Schneider:2012ez,Hoferichter:2012pm,Hoferichter:2014vra,Hoferichter:2018dmo,Hoferichter:2018kwz}, which was mainly motivated by the pion-pole contribution to hadronic light-by-light scattering, but later applied also to specific channels in hadronic vacuum polarization~\cite{Hoferichter:2019gzf,Hoid:2020xjs,Hoferichter:2022iqe,Hoferichter:2023bjm,Hoferichter:2023sli} and, most importantly in the context of $K_L\to \ell^+\ell^-$, to an improved SM prediction for $\pi^0\to e^+e^-$~\cite{Hoferichter:2021lct}. In addition, the calculation presented here profits from the experience gained with the analog form factors in $\eta, \eta'\to\gamma^*\gamma^*$~\cite{Hanhart:2013vba,Holz:2015tcg,Kubis:2015sga,Holz:2022hwz} and $f_1\to \gamma^*\gamma^*$~\cite{Hoferichter:2020lap,Zanke:2021wiq,Hoferichter:2023tgp}. 

The outline of our analysis is as follows. We first derive a dispersive representation of $K_L\to\gamma^*\gamma^*$ in Sec.~\ref{sec:disp}, which is then matched to asymptotic constraints in Sec.~\ref{sec:matching} to arrive at the final representation given in  Sec.~\ref{sec:final}. 
The resulting SM prediction for $K_L\to\ell^+\ell^-$ is derived in Sec.~\ref{sec:KL}, leading to the BSM constraints discussed in Sec.~\ref{sec:BSM}, before concluding in Sec.~\ref{sec:summary}. Details of the SD contribution in the SM, the dispersive representation of the loop integrals, and the integration kernels for $K_L\to\ell^+\ell^-$ are discussed in the appendices. 

\section{Dispersive calculation of $\boldsymbol{K_L\to \gamma^*\gamma^*}$}
\label{sec:disp}

\subsection{Leptonic processes}
\label{sec:leptonic}

Following the conventions of Ref.~\cite{Cirigliano:2011ny}, the amplitude for $K_L\to\gamma^*\gamma^*$ can be written as
\beq
\A^{\mu\nu}[K_L\to\gamma^*(q_1,\mu)\gamma^*(q_2,\nu)]=i\eps^{\mu\nu\alpha\beta}q_{1\alpha}q_{2\beta}c(q_1^2,q_2^2),
\eeq
since the other two Lorentz structures, associated with scalar amplitudes $a(q_1^2,q_2^2)$, $b(q_1^2,q_2^2)$, only lead to $CP$-violating contributions. Accordingly, the scalar function $c(q_1^2,q_2^2)$ defines the $K_L\to \gamma^*\gamma^*$ transition form factor of primary interest to us. Its normalization is determined via the on-shell process
\beq
\label{nor}
\Gamma_{\gamma\gamma}\equiv\Gamma[K_L\to\gamma\gamma]=\frac{\mk^3}{64\pi}\big|c(0,0)\big|^2.
\eeq

The spectrum for the singly-virtual decays $K_L\to\ell^+\ell^-\gamma$ reads
\beq
\frac{1}{\Gamma_{\gamma\gamma}}\frac{d\Gamma_\ell}{dz}=\frac{2}{z}(1-z)^3\frac{1}{\pi}\Im \Pi_\ell(z) |f(z)|^2,
\eeq
with $z=q^2/\mk^2$, $q^2$ the virtuality of the dilepton pair, $f(z)=c(q^2,0)/c(0,0)$, and spectral function
\beq
\frac{1}{\pi}\Im \Pi_\ell(z)=\theta\big(z-4r_\ell^2\big)\frac{\alpha}{3\pi}\bigg(1+2\frac{r_\ell^2}{z}\bigg)\sqrt{1-\frac{4r_\ell^2}{z}}.
\eeq
Similarly, the doubly-virtual process $K_L\to \ell_1^+\ell_1^-\ell_2^+\ell_2^-$ is sensitive to the normalized form factor $f(z_1,z_2)=c(q_1^2,q_2^2)/c(0,0)$, according to
\beq
\frac{1}{\Gamma_{\gamma\gamma}}\frac{d^2\Gamma_{\ell_1\ell_2}}{dz_1dz_2}
=\frac{2}{z_1z_2}\lambda^{3/2}(1,z_1,z_2)
\frac{1}{\pi}\Im \Pi_{\ell_1}(z_1)\frac{1}{\pi}\Im \Pi_{\ell_2}(z_2)|f(z_1,z_2)|^2,
\eeq
where $\lambda(a,b,c)=a^2+b^2+c^2-2(ab+ac+bc)$.
However, because of the kinematic constraints and statistics of these decays, both processes can, in practice, only provide useful information about the slope parameter $b_{K_L}$ defined as 
\beq
\frac{c(q^2,0)}{c(0,0)}=1+b_{K_L}q^2+\Order(q^4).
\eeq

Among the two popular parameterizations of the form factor, Bergstr{\"o}m, Mass{\'o}, and Singer (BMS) have proposed~\cite{Sarraga:1971ohx, Bergstrom:1983rj}
\beq
f_{\text{BMS}}(z)= \frac{M_{\rho}^{2}}{M_{\rho}^{2}-z M_{K}^{2}}+\alpha_{K^{*}}A_{K^{*}}(z), 
\eeq
where
\beq
\label{K*_sl}
A_{K^{*}}(z)=\frac{C_{K^*}M_{K^{*}}^{2}}{M_{K^{*}}^{2}-z M_{K}^{2}}\bigg[\bigg(1+\frac{\beta_{K^*}}{3}\bigg)-\frac{M_{\rho}^{2}}{M_{\rho}^{2}-z M_{K}^{2}}-\frac{\beta_{K^*}}{3}\bigg(\frac{1}{3}\frac{M_{\omega}^{2}}{M_{\omega}^{2}-z M_{K}^{2}}+\frac{2}{3}\frac{M_{\phi}^{2}}{M_{\phi}^{2}-z {M_{K}^{2}}}\bigg) \bigg], 
\eeq
with $C_{K^*}=2.5$ and $\beta_{K^*}=1$ for the resulting  BMS model~\cite{Bergstrom:1983rj}.
The form in Eq.~\eqref{K*_sl} is constructed in such a way that $A_{K^*}(0)=0$, which reflects the fact that the decay $K\to\pi\gamma$ is not allowed by angular momentum conservation, and thus an on-shell $K^*\simeq K\pi\to\gamma$ transition cannot occur either. Phenomenologically, this is an important feature of the $K^*$ contribution, as it can be sizable for the slope of the form factor, without altering the normalization.  
The singly-virtual decays constrain $\alpha_{K^{*}}=-0.217(34)$ (dominated by Ref.~\cite{KTeV:2007ksh})  and $\alpha_{K^{*}}=-0.158(27)$ (dominated by Ref.~\cite{KTeV:2001sfq}),  for the electron and muon channel, respectively.  There is also a measurement from the doubly-virtual process where $\ell_1=\ell_2=e$~\cite{KTeV:2001nui}, which finds  $\alpha_{K^{*}}=-0.14(22)$ with a large uncertainty after assuming factorization for the form factor. The world average,  $\alpha_{K^{*}}=-0.205(22)$~\cite{ParticleDataGroup:2022pth}, then leads to
\beq
\label{slope}
b_{K_L}=2.72(11)\GeV^{-2}.
\eeq
Second, D'Ambrosio, Isidori, and Portol\'es (DIP) have proposed the parameterization~\cite{DAmbrosio:1997eof}
\beq
f(z_1,z_2)=1+\alpha_{\text{DIP}} \sum_{i=1,2}\frac{z_i}{z_i-M_\rho^2/\mk^2}+\beta_{\text{DIP}} \prod_{i=1,2}\frac{z_i}{z_i-M_\rho^2/\mk^2},
\eeq
 often subject to the high-energy constraint $1+2\alpha_{\text{DIP}}+\beta_{\text{DIP}}=0$.
 In addition to the singly-virtual modes~\cite{KTeV:2001sfq, KTeV:2007ksh}, the parameter $\alpha_{\text{DIP}}$ was also extracted from the $\ell_1=e$, $\ell_2=\mu$ channel~\cite{KTeV:2002kut}  assuming $\beta_{\text{DIP}}=0$. The global average, $\alpha_{\text{DIP}}=-1.69(8)$~\cite{ParticleDataGroup:2022pth} is dominated by the $K_L\to e^+e^-\gamma$ mode, leading to  $b_{K_L}=2.81(13)\GeV^{-2}$, consistent with the extraction~\eqref{slope} from the BMS model within uncertainties. 

Our aim is to improve model parameterizations of $c(q_1^2,q_2^2)$,  by developing a dispersive representation for this form factor in analogy to Refs.~\cite{Hoferichter:2018dmo,Hoferichter:2018kwz}. In particular, this allows us to profit not only from the leptonic decay data, but also from data on the related hadronic process $K_L\to\pi^+\pi^-\gamma$, to which we turn next.

\subsection[$K_L\to\pi^+\pi^-\gamma^*$]{$\boldsymbol{K_L\to\pi^+\pi^-\gamma^*}$}
\label{subsec:pipig}

We decompose the $K_L\to\pi^+\pi^-\gamma^*$ amplitude as~\cite{Ecker:1993cq,DAmbrosio:1997hlp,Cirigliano:2011ny}
\beq
\A^\mu[K_L(p_K)\to \pi^+(p_1)\pi^-(p_2)\gamma^*(\mu,q)]
=\frac{E(z_i)}{\mk}\big(z_1 p_2^\mu-z_2 p_1^\mu\big)+\frac{M(z_i)}{\mk^3}i\epsilon^{\mu\alpha\beta\gamma}p_{1\alpha}p_{2\beta}q_\gamma,
\eeq
where $z_i=q\cdot p_i/\mk^2$ and $z_3=z_1+z_2=E_\gamma^*/\mk$, $E_\gamma^*$ denoting the photon energy in the kaon rest frame. As a first step, we rewrite these conventions in terms of Mandelstam variables~\cite{Holz:2015tcg,Kubis:2015sga}
\beq
s=(p_K-p_1)^2=(q+p_2)^2,\qquad t=(p_K-q)^2=(p_1+p_2)^2,\qquad u=(p_K-p_2)^2=(q+p_1)^2,
\eeq
with $s+t+u=2\mpi^2+\mk^2+q^2$,
and the center-of-mass angle
\beq
\cos \theta_t=z_t=\frac{s-u}{\sigma_\pi(t)\lambda^{1/2}(\mk^2,t,q^2)},\qquad \sigma_\pi(t)=\sqrt{1-\frac{4\mpi^2}{t}},
\eeq
leading to the identifications
\beq
z_1=\frac{u-q^2-\mpi^2}{2\mk^2},\qquad z_2=\frac{s-q^2-\mpi^2}{2\mk^2},\qquad 
z_3=\frac{\mk^2-t-q^2}{2\mk^2}=\frac{E_\gamma^*}{\mk}.
\eeq
Moreover, the difference of the $z_i$ is related to $z_t$ according to
\beq
\label{zt}
z_t=\frac{2\mk^2(z_1-z_2)}{\sigma_\pi(t)\lambda^{1/2}(\mk^2,t,q^2)}.
\eeq
The two amplitudes are typically decomposed into ``inner-bremsstrahlung'' (IB) and ``direct emission'' (DE), where the former only contributes to the electric amplitude $E(z_i)$
\beq
E_\text{IB}(z_i)=\frac{e}{z_1z_2\mk}\A[K_L\to\pi\pi]\overset{q^2\to 0}{\to}\frac{e}{1-\sigma_\pi^2(t)z_t^2}\bigg(\frac{2\mk}{E_\gamma^*}\bigg)^2\frac{\A[K_L\to\pi\pi]}{\mk},
\eeq
which is thus  $CP$ violating, but still significant due to the infrared enhancement. The remaining contribution can be expanded in terms of multipoles
\begin{align}
	E_\text{DE}(z_i)&=E_1(z_3)+E_2(z_3) (z_1-z_2)+\Order\big((z_1-z_2)^2\big),\notag\\
	M(z_i)&=M_1(z_3)+M_2(z_3) (z_1-z_2)+\Order\big((z_1-z_2)^2\big),
\end{align}
where the odd electric and even magnetic multipoles are $CP$ violating. This can be contrasted with the expected partial-wave expansion~\cite{Jacob:1959at}, e.g.,
\beq
M(z_i)=\sum_{\text{odd }\ell}P_\ell'(z_t) M_\ell(t),
\eeq
with derivatives of the Legendre polynomials $P_\ell'(z)$,
so that requiring $CP$ invariance indeed only leaves the odd partial waves in $M(z_i)$ (and vice versa for $E(z_i)$). To constrain the $CP$-conserving $K_L\to\gamma^*\gamma^*$ process via the $\pi\pi$ singularities, we are thus mainly interested in the first multipole of $M(z_i)$, which takes the role of the partial wave $f_1(t)$ in Refs.~\cite{Holz:2015tcg,Kubis:2015sga}. 

Experimentally, $M_1(z_3)$ was studied in Refs.~\cite{E731:1992nnl,KTeV:2000avq,KTeV:2006diq} for $q^2=0$, using a $\rho$-pole ansatz of the form~\cite{Lin:1987de}
\beq
\label{M1_pole}
M=\bar M_1\bigg(1+\frac{a}{M_\rho^2-\mk^2+2\mk^2 z_3}\bigg)
=\bar M_1\bigg(1+\frac{a}{M_\rho^2-t}\bigg),\qquad \bar M_1=\frac{e|f_S|}{\mk}\tilde g_{M_1},
\eeq
with $|\tilde g_{M_1}|=1.20(9)$, $a=-0.738(19)\GeV^2$~\cite{KTeV:2006diq}, and the rest of the normalization was expressed in terms of the $K_S\to\pi^+\pi^-$ decay~\cite{Sehgal:1992wm,Sehgal:1999vg}\footnote{This assignment implies the relation $f_S=\A[K_S\to\pi^+\pi^-]$. Assuming the $\Delta I =\frac{1}{2}$ rule, the tree-level chiral prediction gives $|f_S|=2|G_8|F_0(\mk^2-\mpi^2)$. With $|G_8|\approx 9.0\times10^{-6}\,\GeV^{-2}$~\cite{Cirigliano:2011ny}, its value is consistent with the result extracted from the  decay width of $K_S\to\pi^+\pi^-$.} 
\beq
\Gamma[K_S\to\pi^+\pi^-]=\frac{|f_S|^2}{16\pi\mk}\sigma_\pi(\mk^2).
\eeq
Assuming no interference with the IB contribution, a  branching fraction of $\Br[K_L\to\pi^+\pi^-\gamma\text{ (DE)}]=2.84(11)\times 10^{-5}$ is given in Ref.~\cite{ParticleDataGroup:2022pth} (based on the $K_L\to\pi^+\pi^-$ measurement from Ref.~\cite{KLOE:2006ane} and the ratio of IB vs.\ DE branching fractions from Refs.~\cite{KTeV:2000avq,KTeV:2006diq,E731:1992nnl}).
Finally, there is a measurement of the branching fraction $\Br[K_L\to\pi^+\pi^- e^+e^-]=3.08(20)\times 10^{-7}$~\cite{NA48:2003pwz}. Apart from branching fractions, experiments also performed analyses of the double differential decay rate 
\begin{align} 
	\frac{ \diff \Gamma[K_L\to\pi^+\pi^-\gamma ]}{ \diff E_\gamma^* \,\diff \cos \theta}=& \left(\frac{\beta E_\gamma^*}{8 \pi M_{K}}\right)^{3}\left(1-\frac{2 E_\gamma^*}{M_{K}}\right) \sin ^{2} \theta\big(|E_\text{IB}+E_{\text {DE}}|^{2}+|M|^{2}\big), 
\end{align}
with $\beta=\big(1-4 M_{\pi}^{2} /(M_{K}^{2}-2 M_{K}E_\gamma^*)\big)^{1 / 2}$, which facilitate spectral-shape studies of the magnetic amplitude $M$.  In this regard,  a representation such as Eq.~\eqref{M1_pole} cannot be justified from a dispersive perspective, since the sum of a contact term and a $\rho$ pole violates Watson's final-state theorem~\cite{Watson:1954uc}.  To improve upon existing parameterizations the next step thus consists of restoring unitarity as a minimal requirement.

\subsection{Unitarity and dispersion relations}

\begin{figure}[t]
	\centering
	\includegraphics[width=.9\linewidth]{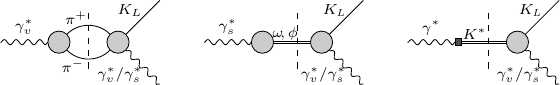}
	\caption{Different contributions to the $K_L\to\gamma^*\gamma^*$ form factor in a dispersive approach. The gray blobs denote the respective hadronic matrix elements, and the black square the strangeness-changing $K^*$--$\gamma^*$ coupling that is again mediated by vector-meson states. } 
	\label{fig:FF}
\end{figure}

As a first step, we decompose $c(q_1^2,q_2^2)$ into functions with definite isospin and a term that collects the $K^*$ contribution as detailed in Fig.~\ref{fig:FF}: 
\beq
c^\text{disp}(q_1^2,q_2^2)=c_{vv}(q_1^2,q_2^2)+c_{vs}(q_1^2,q_2^2)+c_{sv}(q_1^2,q_2^2)+c_{ss}(q_1^2,q_2^2)+c^\text{disp}_{K^*}(q_1^2,q_2^2),
\eeq
where 
\beq
c^\text{disp}_{K^*}(q_1^2,q_2^2)=c_{vK^*}(q_1^2,q_2^2)+c_{K^*v}(q_1^2,q_2^2)+c_{sK^*}(q_1^2,q_2^2)+c_{K^*s}(q_1^2,q_2^2), 
\eeq
and the two photons have isovector ($v$) and isoscalar ($s$) quantum numbers as indicated by the subscripts.  Bose symmetry implies $c_{vv}(q_1^2,q_2^2)$ and $c_{ss}(q_1^2,q_2^2)$ to be symmetric under the exchange of the arguments, and for the mixed terms $c_{vs}(q_1^2,q_2^2)=c_{sv}(q_2^2,q_1^2)$, etc. Phenomenologically, the necessity of the $K^*$ contribution follows from the large slope parameter~\eqref{slope}, $b_{K_L}=2.72(11)\GeV^{-2}\gg M_\rho^{-2}\simeq 1.66\GeV^{-2}$,  as opposed to those in $\pi^0\to\gamma^*\gamma$~\cite{Hoferichter:2014vra,Hoferichter:2018dmo,Hoferichter:2018kwz} and $\eta, \eta'\to\gamma^*\gamma$~\cite{Hanhart:2013vba,Holz:2015tcg,Kubis:2015sga,Holz:2022hwz} not far from the typical scale $M_\rho^{-2}$.

In the same vain,  $M(t,q^2)=M_1(z_3)/\mk^3$, the (normalized) multipole for general photon virtuality $q^2$,  is decomposed as 
\beq
\label{decomp}
M(t,q^2)=M_v(t,q^2)+M_s(t,q^2)+M_{K^*}(t,q^2).
\eeq  
The two-pion cuts then lead to the unsubtracted dispersion relations
\begin{align}
	\label{disp_ansatz}
	c_{vv}(q_1^2,q_2^2)&=\frac{e}{12\pi^2}\int_{4\mpi^2}^{\sm} \diff x\frac{q_\pi^3(x)[F_\pi^V(x)]^* M_v(x,q_2^2)}{x^{1/2}(x-q_1^2)}\notag\\
	&=\frac{e}{12\pi^2}\int_{4\mpi^2}^{\sm}  \diff x\frac{q_\pi^3(x)[F_\pi^V(x)]^* M_v(x,q_1^2)}{x^{1/2}(x-q_2^2)},\notag\\
	c_{vs}(q_1^2,q_2^2)&=\frac{e}{12\pi^2}\int_{4\mpi^2}^{\sm}  \diff x\frac{q_\pi^3(x)[F_\pi^V(x)]^* M_s(x,q_2^2)}{x^{1/2}(x-q_1^2)},\notag\\
	c_{sv}(q_1^2,q_2^2)&=\frac{e}{12\pi^2}\int_{4\mpi^2}^{\sm}  \diff x\frac{q_\pi^3(x)[F_\pi^V(x)]^* M_s(x,q_1^2)}{x^{1/2}(x-q_2^2)},
\end{align}
where  $q_\pi(x)=\sqrt{x/4-\mpi^2}$, $F_\pi^V$ is the electromagnetic form factor of the pion, and $\sqrt{\sm}=2 \GeV$ denotes a cutoff that separates the low-energy contribution.   

\subsubsection{Factorization}

The simplest solution to the decomposition~\eqref{decomp} involves separating the final-state interaction into  an Omn\`es factor~\cite{Omnes:1958hv} 
\beq
\label{Omnes}
\Omega_1(t)=\exp\bigg\{\frac{t}{\pi}\int_{4\mpi^2}^\infty \diff x\frac{\delta_1(x)}{x(x-t)}\bigg\},
\eeq
determined by the $P$-wave $\pi\pi$ scattering phase shift $\delta_1(t)$, and parameterizing the remainder via polynomials
\beq
\label{factorization}
\M(t, q^{2})=\sum_{i=v, s} a_{i}(q^{2})\bigg[P_{i}(t)+\epsilon_{K^*}\frac{ a_{K^{*}}(t)}{ a_{K^{*}}(0)}\bigg] \Omega_1(t),
\eeq
with $P_{v/s}(t)$ for the isovector/isoscalar contribution, a parameter $\epsilon _{K^*}$ for the $K^*$, 
and another function $a_{v/s}(q^2)$ to describe the dependence on the photon virtuality. In addition,
\beq\label{eq:ak}
a_{K^{*}}(q^2)=\frac{1}{\pi}\int_{(\mk+M_\pi)^2}^{\sm} \diff x \frac{ \Im \text{BW}_{K^{*}}(x)}{x-q^2}, 
\eeq
where the imaginary part follows a  Breit--Wigner description of the $K^*$ resonance.\footnote{This approximation ensures the absence of unphysical imaginary parts below threshold~\cite{Lomon:2012pn,Moussallam:2013una,Crivellin:2022gfu}, and could be further improved by an Omn\`es factor for $\pi K$ scattering in analogy to Eq.~\eqref{Omnes}, but in view of the remaining uncertainties discussed below, e.g., from the coupling $\eps_{K^*}$,  such refinements are currently not mandated. Our treatment of the $K^*$ contribution essentially follows the implementation of $\rho$--$\omega$ mixing in Ref.~\cite{Hoferichter:2016duk}.}  Using the ansatz~\eqref{factorization}, the $c_{vv}$ contribution in 
Eq.~\eqref{disp_ansatz} can be rewritten as
\begin{align}
\label{cvv_factorization}
c_{vv}(q_1^2,q_2^2)&=a_{v}(q_1^2)a_{vv}(q_2^2)=a_{v}(q_2^2)a_{vv}(q_1^2),\notag\\
a_{vv}(q^2)&=\frac{1}{\pi}\int_{4\mpi^2}^{\sm} \diff x\frac{\rho_{vv}(x)}{x-q^2},\qquad \rho_{vv}(x)=\frac{e q_\pi^3(x)[F_\pi^V(x)]^* P_v(x)\Omega_1(x)}{12\pi \sqrt{x}}. 
\end{align}
The first line implies the factorization $a_{vv}(q^2)=\alpha_1 a_v(q^2)$. Moreover, Eq.~\eqref{disp_ansatz} remains invariant upon rescaling $a_v(q^2)\to\alpha_2 a_v(q^2)$, $P_v(t)\to P_v(t)/\alpha_2$, i.e., $a_{vv}(q^2)=\alpha_1\alpha_2^2 a_v(q^2)$. Choosing $\alpha_2=1/\sqrt{|\alpha_1|}$ therefore shows that 
\beq
\label{cvv_factorized}
c_{vv}(q_1^2,q_2^2)=\pm a_{vv}(q_1^2)a_{vv}(q_2^2).
\eeq
It is further instructive to consider the limit of a narrow resonance~\cite{Hoferichter:2012pm}
\beq
\label{a_v_narrow}
a_{vv}(q^2)\to e F_\pi^2 P_v(M_\rho^2)\frac{M_\rho^2}{M_\rho^2-q^2},
\eeq
where the KFSR relation~\cite{Kawarabayashi:1966kd,Riazuddin:1966sw} has been used to trade the $\rho\pi\pi$ coupling in favor of the pion decay constant $F_\pi$. Up to a normalization, $a_{vv}(q^2)$ thus reduces to a $\rho$ propagator in the narrow-resonance approximation.

The isoscalar function $a_s(q^2)$ is dominated by $\omega$ and $\phi$ resonances, leading to the ansatz 
\beq
\label{as_BW}
a_s(q^2)=\sum_{V=\omega,\phi}c_V  \frac{a_V(q^2)}{a_V(0)}, \qquad a_V(q^2)=\frac{1}{\pi}\int_{\sth}^{\sm} \diff x  \frac{\Im \text{BW}_V(x)}{x-q^2}, 
\eeq
with a spectral function determined by a suitable parameterization of the imaginary parts following~\cite{Hoferichter:2014vra,Hoferichter:2018dmo,Hoferichter:2018kwz}.\footnote{The lowest threshold in the isospin limit is $\sth=9\mpi^2$, while isospin-breaking corrections from $\omega\to\pi^0\gamma$ can induce a contribution starting at $\sth=\mpi^2$.} In the limit of a narrow resonance, this form becomes identical to Eq.~\eqref{a_v_narrow} upon identifying $c_\rho=e F_\pi^2 P_v(M_\rho^2)$. For the mixed isovector--isoscalar contribution we have
\begin{align}
c_{vs}(q_1^2,q_2^2)&=a_s(q_2^2)a_{vs}(q_1^2),\notag\\
a_{vs}(q^2)&=\frac{1}{\pi}\int_{4\mpi^2}^{\sm} \diff x\frac{\rho_{vs}(x)}{x-q^2},\qquad \rho_{vs}(x)=\frac{e q_\pi^3(x)[F_\pi^V(x)]^* P_s(x)\Omega_1(x)}{12\pi \sqrt{x}},
\end{align}
where the difference to $a_{vv}(q^2)$ solely originates from the polynomial $P_{v/s}(t)$. 
 Finally, for the isoscalar--isoscalar contribution we set
\beq
c_{ss}(q_1^2,q_2^2)= \sum_{V,V'=\omega,\phi} c_{VV'}\frac{a_V(q_1^2)}{a_V(0)}\frac{a_{V'}(q_2^2)}{a_{V'}(0)}, 
\eeq
in analogy to Eq.~\eqref{as_BW}. For the actual application, we further need to determine the overall sign and the relative size of the various isospin components, a point to which we will return in Sec.~\ref{sec:VMD}.

The remaining parts enter in the context of  the $K^*$ contribution,
the third diagram in Fig.~\ref{fig:FF}. To evaluate this diagram, we analyze the double discontinuity arising from the two-pion and $\pi K\simeq K^*$ intermediate states, in analogy to $2\pi$ and $3\pi$ singularities in the context of $\rho$--$\omega$ mixing in Ref.~\cite{Hoferichter:2016duk}. We obtain the result
\begin{align}
\label{Kstar_parameterization}
	c_{vK^*}(q_1^2,q_2^2)&=\pm a_{vv}(q_1^2)\epsilon_{K^*}\frac{ a_{K^{*}}(q_2^2)}{ a_{K^{*}}(0)} 2a_\rho(0)\notag\\
	&\times\Re\bigg[\bigg(1+\frac{\beta_{K^{*}}}{3}\bigg)-\frac{ a_{\rho}(q_2^{2})}{ a_\rho(0)}-\frac{\beta_{K^{*}}}{3}\bigg(\frac{1}{3}\frac{ a_{\omega}(q_2^2)}{ a_{\omega}(0)}+\frac{2}{3}\frac{ a_{\phi}(q_2^{2})}{ a_{\phi}(0)}\bigg) \bigg],
\end{align}
where the overall sign is related to the one in $c_{vv}(q_1^2,q_2^2)$
and the same expression holds for $c_{sK^*}(q_1^2,q_2^2)$ with $a_{vv}(q_1^2)\to a_{s}(q_1^2)$. The contribution from the $\rho$ can be derived from the double discontinuity, leading to  
\beq
 a_{\rho}(q^{2})= \frac{1}{\pi}\int_{4\mpi^2}^{\sm} \diff x\frac{\rho_\rho(x)}{x-q^2},\qquad \rho_\rho(x)=\frac{eq_\pi^3(x)[F_\pi^V(x)]^* \Omega_1(x)}{12\pi\sqrt{x}}, 
\eeq
which coincides with $a_{vv}(q^2)$, $a_{vs}(q^2)$ in the limit $P_{v/s}(t)=1$. The coefficients of the $\omega$ and $\phi$ contributions are parameterized in analogy to the BMS ansatz~\eqref{K*_sl}, with the propagators replaced by a dispersively improved Breit--Wigner prescription as defined in Eq.~\eqref{as_BW}. The remaining free parameters are $\epsilon_{K^*}$ and  $\beta_{K^{*}}$, where the latter is necessary to accommodate the slope parameter~\eqref{slope}.  

Factorization-breaking contributions to $M_{v/s}(t,q^2)$ are generated when including left-hand cuts (LHCs), which are expanded in a polynomial in Eq.~\eqref{factorization}. Neglecting isospin-breaking contributions, the leading LHCs to $K_L\to\pi^+\pi^-\gamma^*$ arise from a pion pole (for $M_{v}(t,q^2)$) and two-pion intermediate states (for $M_{s}(t,q^2)$). Since the former involves a $K_L\to\pi\pi$ vertex, this contribution will be $CP$ violating. The two-pion LHC in $M_{s}(t,q^2)$ conserves $CP$, but involves an anomalous $\gamma3\pi$ vertex (and only appears in the isoscalar amplitude). Accordingly, the role of LHCs beyond their polynomial approximation should be negligible, especially in the limited kinematic domain accessible in the kaon decay.  

\subsubsection[Fits to $K_L\to\pi^+\pi^-\gamma$]{Fits to $\boldsymbol{K_L\to\pi^+\pi^-\gamma}$}

As anticipated in Sec.~\ref{subsec:pipig}, experiments have observed evidence for an $E_\gamma^*$-dependent form-factor modification in contrast to a  pure-$M_1$ DE amplitude~\cite{E731:1992nnl,KTeV:2000avq,KTeV:2006diq}. This enables us to determine the polynomial in Eq.~\eqref{factorization}. Accordingly, we perform a linear-polynomial fit to the spectrum data from Refs.~\cite{KTeV:2000avq} by
\beq
\label{polynomials}
P_{v/s}(t)=N_{v/s}(1+\alpha_{v/s}t). 
\eeq
In the singly-virtual decay, it is not possible to separate all isospin components, instead, the spectrum is sensitive to the parameter combinations
\begin{align}
\label{par_sum}
N_{\pi\pi}&=a_{vv}(0)N_v+a_s(0)N_s, \qquad b_{\pi\pi}=a_{vv}(0)N_v\alpha_{v}+a_s(0)N_s\alpha_{s}, \notag\\ 
\eta_{K^*}&= (a_{vv}(0)+a_s(0))\epsilon_{K^*}.
\end{align}
\begin{table*}[t]
	\centering
	\renewcommand{\arraystretch}{1.3}
	\small
	\begin{tabular}{l l l l l}
		\toprule
		 $N_{\pi\pi} [\GeV^{-3}]$&$\bar b_{\pi\pi} [\GeV^{-2}]$ &$\bar \eta_{K^*}$& $\chi^2/\text{dof}$ & $p$-value\\\midrule
		$ 3.47(27)\times 10^{-8}$  & $10.0(1.9)$ & $0.96(10)$   &$34.1/27=1.26$& $16.3\%$ \\
		\bottomrule
		\renewcommand{\arraystretch}{1.0}
	\end{tabular}
	\caption{Fit to the $K_L\to\pi^+\pi^-\gamma$ spectrum from Ref.~\cite{KTeV:2000avq}, see main texts for more details.}
	\label{tab:fit}
\end{table*}

In particular, it is useful to rewrite the amplitude as
\beq
\label{KLpipigamma_fit}
N_{\pi\pi}\bigg[1+\bar b_{\pi\pi}t+\bar\eta_{K^*}\frac{a_{K^*}(t)}{a_{K^*}(0)}\bigg],
\eeq
with
\beq
b_{\pi\pi}=N_{\pi\pi}\bar b_{\pi\pi},\qquad \eta_{K^*}=N_{\pi\pi}\bar \eta_{K^*},
\eeq
to isolate normalization and $t$ dependence in the fit. Ideally,  one should then perform a combined analysis of all data for the spectra of the leptonic decays $K_L\to\ell^+\ell^-\gamma$ and the  hadronic channel $K_L\to\pi^+\pi^-\gamma$, in order to disentangle the linear term in Eq.~\eqref{KLpipigamma_fit} from the $K^*$ contribution. Unfortunately, the original data for these spectra are rarely available: the best measurements of $K_L\to e^+e^-\gamma$~\cite{KTeV:2007ksh} and $K_L\to\mu^+\mu^-\gamma$~\cite{KTeV:2001sfq} are only provided in terms of fits using the two models introduced in Sec.~\ref{sec:leptonic}, but the original data for the spectrum are not included in the publications. Similarly, we could only obtain the $K_L\to\pi^+\pi^-\gamma$ data from Ref.~\cite{KTeV:2000avq} by digitizing Fig.~4 therein (we checked that a fit to the digitized data reproduces the fit parameters given in the paper), while  the spectra from Refs.~\cite{E731:1992nnl,KTeV:2006diq} are not accessible at all.

\begin{figure}[t]
	\centering
	\includegraphics[width=.8\linewidth]{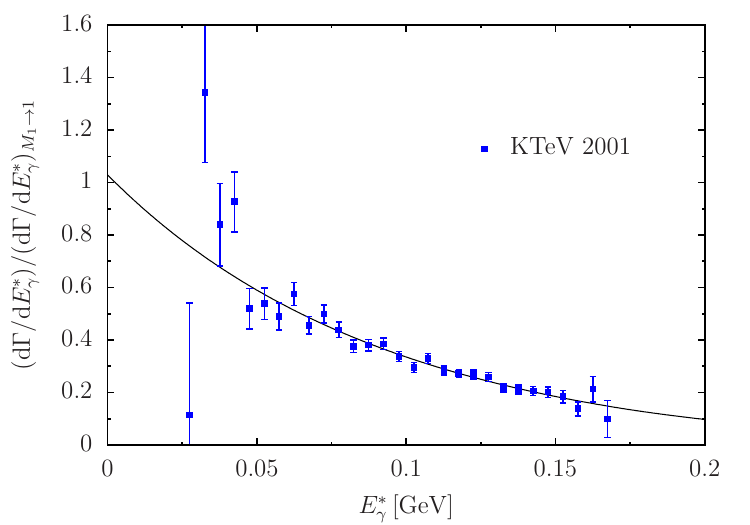}
	\caption{Fit to the  $K_L\to\pi^+\pi^-\gamma$ differential decay rate, using the data from Ref.~\cite{KTeV:2000avq}. Vertical scale in arbitrary units.} 
	\label{fig:M1}
\end{figure}

Given the high degree of correlation between the linear polynomial and the $K^*$ piece, the constraints from the dilepton data are critical, in such a way that we do need to rely on the value of $\eta_{K^*}$ inferred from the global average of $\alpha_{K^{*}}$~\cite{ParticleDataGroup:2022pth}. Accordingly, we will also use the slope determination via the 
BMS parameterization~\eqref{slope},
to ensure a consistent set of input quantities. The resulting residual model dependence could be avoided if the spectral data from Refs.~\cite{KTeV:2007ksh,KTeV:2001sfq,E731:1992nnl,KTeV:2006diq} were available. With $\eta_{K^*}$ constrained to its BMS value, we then  fit the remaining parameters $N_{\pi\pi}$ and $\bar b_{\pi\pi}$  to the spectrum from Ref.~\cite{KTeV:2000avq}, first, with normalization in arbitrary units. 
In a final step, $N_{\pi\pi}$ is rescaled to reproduce the total DE decay rate~\cite{ParticleDataGroup:2022pth},  for which our fit produces an uncertainty estimate consistent with Ref.~\cite{KTeV:2000avq} after taking into account the correlations between the fit parameters. The fit results are shown in Table~\ref{tab:fit} and Fig.~\ref{fig:M1}, and will serve as basis for the error estimate of the dispersive form factor in Sec.~\ref{sec:KL}.\footnote{We use the phase-shift input from Ref.~\cite{Colangelo:2018mtw}, but the effects from the phase shift, e.g., in comparison to the earlier determinations from Ref.~\cite{Caprini:2011ky, Garcia-Martin:2011iqs}, are negligible compared to the other sources of uncertainty of the dispersive form factor.} 

\subsection{Vector meson dominance}
\label{sec:VMD}

The $K_L\to\pi^+\pi^-\gamma$  channel only determines the sum of the isovector and isoscalar parameters in the form of Eq.~\eqref{par_sum} as they enter in the singly-virtual form factor,  but not the relative weights between the different isospin contributions.  In such a situation,  the vector-meson-dominance (VMD) picture often provides useful information to constrain the relative strength of different isospin components in the doubly-virtual form factor as well, see, e.g., Refs.~\cite{Klingl:1996by,Gan:2020aco,Zanke:2021wiq}. 

The $SU(3)$ pseudoscalar matrix $|\phi\rangle$ transforms under $CP$ as $CP|\phi\rangle=-|\phi\rangle^T$. The phase in front is fixed by the $CP$ quantum numbers of the neutral states, e.g., from $CP|\pi^0\rangle=-|\pi^0\rangle$. Thereafter, the $K_L$ state is defined as $(|K^0 \rangle+|\bar {K^0}\rangle)/\sqrt{2}$, in line with the prescription applied in Ref.~\cite{Ecker:1991ru}. 
After identifying the phase convention in this way, we can write down the simplest trace that correctly projects out the $K_L$ state, 
\beq
\text{Tr}\big(\{\lambda_6,|\phi\rangle\} |V\rangle  |V\rangle\big), 
\eeq
where $|V\rangle$ is the vector-meson nonet matrix and we assume ideal mixing for $\omega$ and $\phi$. Adding the relative factors from the $V\gamma$ couplings~\cite{Klingl:1996by,Hoferichter:2017ftn,Zanke:2021wiq}, the relative weights between the  different contributions to the form-factor normalization read
\beq
\label{nor_ratio}
c_{vv}(0,0): c_{vs}(0,0):c_{ss}(0,0)=1:-\frac{1}{3}:\frac{5}{9}. 
\eeq
The same calculation also predicts the relative size of $\omega$ and $\phi$ contributions to $a_s(q^2)$ and $a_{ss}(q_1^2,q_2^2)$, in particular
\beq
\label{c_isospin}
c_{\phi}=0,\qquad c_{\phi\omega}=c_{\omega\phi}=0,\qquad 
c_{\phi\phi}=4c_{\omega\omega}.
\eeq
Similarly, the trace
\beq
\label{Tr_Kstar}
\text{Tr}\big(\{\lambda_6,|V\rangle\} |V\rangle\big),
\eeq
in combination with the $V\gamma$ couplings,
predicts the relative size of the $K^*$--$\gamma$ transitions
\beq
\label{c_Kstar_VMD}
c_{K^*\to\rho\to\gamma}:c_{K^*\to\omega\to\gamma}:c_{K^*\to\phi\to\gamma}=1:-\frac{1}{3}:\frac{2}{3}.
\eeq
These ratios differ from the ones assumed in the BMS model~\eqref{K*_sl}, $1:1/9:2/9$, which are obtained when removing the singlet from $|V\rangle$ in Eq.~\eqref{Tr_Kstar}, as done in Ref.~\cite{Sarraga:1971ohx}. The two variants give an identical decomposition into isoscalar and isovector contributions, but the relative size of $\omega$ and $\phi$ differs. Although the resulting change in the $K^*$ parameterization~\eqref{Kstar_parameterization} scales with $M_\phi-M_\omega$, we observe that the effect is numerically significant. Accordingly, since we need to rely on $\eps_{K^*}$ from the BMS fits, we employ the same decomposition of $\omega$ and $\phi$, leading to the form anticipated in Eq.~\eqref{Kstar_parameterization}.   
 Finally, while the normalizations $c_{v K^*}(0,0)=c_{s K^*}(0,0)$ vanish by construction, we can consider the ratio
\beq
\frac{c_{v K^*}(q_1^2,q_2^2)}{c_{s K^*}(q_1^2,q_2^2)}=\frac{a_{vv}(q_1^2)}{a_s(q_1^2)},
\eeq
which is constrained in the dispersive approach in terms of $a_{vv}(q^2)$ and $a_s(q^2)$. Here, VMD predicts the same relative weights of $\rho$, $\omega$, and $\phi$ as in Eq.~\eqref{c_Kstar_VMD},
which does not agree with $c_\phi=0$ in $a_s(q^2)$, i.e., in a similar way as in the comparison to the BMS parameterization we observe an ambiguity in the separation of $\omega$ and $\phi$ contributions. The relative normalization
\beq
\label{avv_as_Kstar}
\frac{c_{v K^*}(0,q_2^2)}{c_{s K^*}(0,q_2^2)}=\frac{a_{vv}(0)}{a_s(0)}=3
\eeq
is again not affected, and this constraint 
proves very useful for disentangling the different normalizations, see Sec.~\ref{sec:final}. Beyond, we set
\beq
c_\phi=2c_\omega,
\eeq
for consistency with Eq.~\eqref{Kstar_parameterization} and the BMS input for the $K^*$ piece, while for the non-$K^*$ contributions we keep the (internally consistent) set of conditions~\eqref{nor_ratio}
and~\eqref{c_isospin}.

Ultimately, the impact of these VMD assumptions on the final uncertainty proves relatively minor, for the following reasons: the singly-virtual form factor is probed directly in terms of $K_L\to\ell^+\ell^-\gamma$ and $K_L\to\pi^+\pi^-\gamma$, so that VMD arguments only enter in the doubly-virtual direction. Second, the ambiguities related to $\omega$ and $\phi$ only affect the internal decomposition of the isoscalar contribution, and, e.g., the three variants $c_\phi=\pm2c_\omega$, $c_\phi=0$ for the $K^*$ affect $\Re \A_\mu(\mk^2)$ at a level below $0.1$. In general, we find that the 
result for the reduced amplitude $\Re\A_\ell(\mk^2)$ is rather robust against variation of the relative VMD weights, see Sec.~\ref{sec:KL}, once the constraints from the form factor normalization~\eqref{nor} and slope~\eqref{slope} are imposed. As discussed in Sec.~\ref{sec:final}, only one of the relative normalizations~\eqref{nor_ratio} needs to be provided to determine all parameters, so that choosing either one of them serves as a way to assess the impact of the VMD assumptions, and the corresponding variation will be included in the uncertainty estimate for the dispersive part of the form factor quoted in
Sec.~\ref{sec:KL}.

\section{Matching to asymptotic constraints}
\label{sec:matching}

\begin{figure}[t]
	\centering
	\includegraphics[width=.9\linewidth]{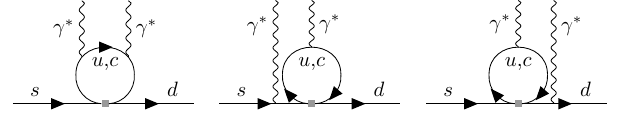}
	\caption{Partonic diagrams for $K_L\to\gamma^*\gamma^*$.} 
	\label{fig:asym}
\end{figure} 

The asymptotic form of $c(q_1^2,q_2^2)$ was constrained in 
Ref.~\cite{Isidori:2003ts} using arguments from a partonic calculation, which we will incorporate here in our dispersive representation. The key result reads
\beq
\label{asym}
c^\text{asym}(q_1^2,q_2^2)=\frac{16\alpha G_F V_{us}^*V_{ud} F_K}{9\pi\sqrt{2}}\big[C_2(\mu)+3C_1(\mu)\big]\bigg[I(q_1^2,q_2^2)+T(q_1^2)+T(q_2^2)\bigg],
\eeq
with tree-level Wilson coefficients normalized as $C_2(M_W)=1$, $C_1(M_W)=0$. The loop functions originate from diagrams with $c$- and $u$-quark loops as depicted in Fig.~\ref{fig:asym}, where either both photons couple in the loop, $I(q_1^2,q_2^2)$, or just one, $T(q_i^2)$, and result from the difference of the two flavors.\footnote{The contribution involving the difference of $t$- and $c$-quarks, proportional to $V_{ts}^*V_{td}$, is negligible compared to the dominant uncertainties from the running of the Wilson coefficients $C_i(\mu)$, see Refs.~\cite{Eeg:1996pr,Isidori:2003ts}.} However, we note that Eq.~\eqref{asym} as in Ref.~\cite{Isidori:2003ts} is valid for deeply-virtual kinematics and up to power-suppressed terms of $\Order(\xi)$, $\xi=m_u^2/m_c^2$.  In order to establish a  matching between the low-energy dispersive contributions and the asymptotic constraints, we start from the full expressions including the dependence on $m_u$, and separate the low- and high-energy contributions by introducing a cutoff in a dispersive representation of the respective loop integrals, see  App.~\ref{app:loop} for details.

\subsection{Two-point function}

The starting point for $T(q^2)$  is given by the usual vacuum polarization from a quark loop, i.e.,
\begin{align}
	\label{T_der}
	T(q^2)&=T_c(q^2)-T_u(q^2)=2\int_0^1  \diff x \,x(1-x)\log\Big[1-x(1-x)\frac{q^2}{m_c^2}\Big] - (c\leftrightarrow u)\notag\\
	&=2\int_0^1  \diff x \,x(1-x)\log\frac{1-x(1-x)r}{\xi-x(1-x)r}+\frac{1}{3}\log \xi,\qquad r=\frac{q^2}{m_c^2}.
\end{align}
It is clear that $T(r)$ does not contribute to the form factor normalization, a property that should be preserved when removing the low-energy contributions. Moreover, for $q^2\to\infty$ the difference of $c$- and $u$-quark integrals decreases as $1/q^2$, which should also be reflected by the final representation. 

The definition of $T(q^2)$ in Ref.~\cite{Isidori:2003ts} follows when dropping the $\log \xi$ term and the dependence on $\xi$ in the integral~\eqref{T_der}, i.e.,
\beq
\label{Tr}
T(r)\big|_\text{\cite{Isidori:2003ts}}=2\int_0^1 \diff x \,x(1-x)\log\frac{1-x(1-x)r}{-x(1-x)r}.
\eeq
This loop function has a good asymptotic behavior for $r\to\infty$
\beq
T(-r_1)+T(-r_2)=\frac{2}{r_1}+\frac{2}{r_2}=\frac{8}{r_1+r_2}\bigg[1+\Order\bigg(\frac{(r_1-r_2)^2}{(r_1+r_2)^2}\bigg)\bigg],
\eeq
but displays a sensitivity to infrared scales.   In particular, $T(r)$ is now divergent for $r\to 0$, as can be made explicit by rewriting Eq.~\eqref{Tr} as
\beq
T(r)\big|_\text{\cite{Isidori:2003ts}}=2\int_0^1  \diff x \,x(1-x)\log\big[1-x(1-x)r\big]
+\frac{5}{9}-\frac{1}{3}\log(-r).
\eeq
For the asymptotic matching, we thus need to impose integration cutoffs in the dispersive representation for $T(q^2)$
\beq
T^\text{asym}(q^2)=T_c^\text{asym}(q^2)-T_u^\text{asym}(q^2)=-\frac{q^2}{3}\bigg[\int_{s_c}^\infty\diff s\frac{\sigma_c(s)(s+2m_c^2)}{s^2(s-q^2)}-\int_{s_u}^\infty \frac{\diff s}{s(s-q^2)}\bigg],
\eeq
where $\sigma_q(s)=\sqrt{1-4m_q^2/s}$,
see Eq.~\eqref{TcTu}. Here, the integration cutoffs $s_u$ and $s_c$ are related to ensure the correct asymptotic behavior, e.g., for the canonical choice $s_c=4m_c^2$ one has $s_u=\mathrm{e}^{5/3}m_c^2\simeq 5.29m_c^2$.

 \subsection{Three-point function}

 As detailed in App.~\ref{app:loop}, the full loop function 
\beq
I(q_1^2,q_2^2)= I_c(q_1^2,q_2^2)- (c\leftrightarrow u)
\eeq
can be reconstructed from Refs.~\cite{Simma:1990nr,Herrlich:1991bq}.  
The resulting expression is  well-defined for all virtualities, thus correctly reproduces the partonic contributions to the real-photon kinematics obtained in Refs.~\cite{Gaillard:1974hs,Gaillard:1975ds,Ma:1981eg}, i.e., it simplifies to
\beq
I_c(0,0)=-1-\frac{2}{r_3}\int_0^1\diff x \frac{\log\big[1-r_3x(1-x) \big]}{x},
\eeq
where $r_3=\mk^2/m_c^2$. We also reproduce the result of Ref.~\cite{Isidori:2003ts}, by taking $r_3=0$ and combining the $u$- and $c$-quark contributions,
\begin{align}
	\label{Iri}
	I(r_1,r_2)\big|_\text{\cite{Isidori:2003ts}}
	&=\frac{1}{6}+\int_0^1 dx\int_0^{1-x}dy\bigg\{\frac{2-x-y+(r_1 x+ r_2 y)(1-x-y)^2}{1-(r_1 x+r_2 y)(1-x-y)}\notag\\
	&-(2-3x-3y)\log\frac{1-(r_1 x+r_2 y)(1-x-y)}{ -(r_1 x+r_2 y)(1-x-y)}\bigg\}+\Order(\xi\log\xi).
\end{align}
 Although the limit $\xi\to 0$ is finite, the final expression in Eq.~\eqref{Iri} is not well-defined for all virtualities, and we again need to isolate the low- and high-energy contributions by imposing a cutoff in the dispersive representation. 
 
 As a first step, we analyzed the discontinuities of $I_c(q_1^2,q_2^2)$, which shows that even for general $r_3$ the expression reduces to a simple scalar loop integral
 \beq
I_q(q_1^2,q_2^2)=-2C_0(r_1,r_2,r_3)-1,
 \eeq
 with $C_0$ as defined in Eq.~\eqref{C0}. Moreover, the analysis in App.~\ref{app:loop} shows that upon introducing a cutoff $s_u$ in the dispersion relation, no infrared singularities occur and the limit $m_u\to 0$ can be taken, in such a way that 
 \beq
I_u(q_1^2,q_2^2)\to -1, 
 \eeq
 i.e., no non-trivial contributions remain from the $u$-quark loop. For the $c$-quark loop, setting $\mk=0$ permits a very simple double-spectral representation of the form
 \begin{align}
I_c(q_1^2,q_2^2)&=-1+\frac{m_c^2}{\pi^2}\int_{s_c}^\infty \diff x\int_{s_c}^\infty \diff y\frac{2\pi^2\delta(x-y)\log\frac{1+\sigma_c(x)}{1-\sigma_c(x)}}{(x-q_1^2)(y-q_2^2)}\notag\\
&=-1+\frac{m_c^2}{\pi}\int_{s_c}^\infty \diff x\frac{2\pi\log\frac{1+\sigma_c(x)}{1-\sigma_c(x)}}{(x-q_1^2)(x-q_2^2)}.
\end{align}
A generalization of the de-facto single-variable dispersion relation in the second line to the case $\mk>0$ is given in Eq.~\eqref{Ic_single}, while a full double-spectral representation requires a contour deformation according to Eq.~\eqref{disp_double_r3_2}. We observed that the numerical difference between the  $\mk=0$ simplified form and the full representation~\eqref{disp_double_r3_2} is negligible for the current application.    

\section{Final representation}
\label{sec:final}

\subsection[Decomposition of $K_L\to\gamma^*\gamma^*$ form factor]{Decomposition of $\boldsymbol{K_L\to\gamma^*\gamma^*}$ form factor}

To derive the final representation we first need to establish the sign convention for the $K_L\to\gamma^*\gamma^*$ form factor $c(q_1^2,q_2^2)$. A detailed discussion of the sign of the SD contribution relative to the LD part is provided in Ref.~\cite{Isidori:2003ts}, under the assumption that the dominant contribution to $\A^{\mu\nu}[K_L\to\gamma\gamma]$ arises from the $\pi^0$ pole
\beq
c(0,0)\big|_{\pi^0}=\frac{2G_8 F_\pi\alpha}{\pi}\frac{\mk^2}{\mk^2-\mpi^2},\qquad \big|c(0,0)\big|_{\pi^0}\simeq 4.2\times 10^{-9} \GeV^{-1},
\eeq
indeed close to the experimental value $|c(0,0)|=3.389(14)\times 10^{-9}\GeV^{-1}$. Accordingly, it seems safe to assume that the signs of $\A[K_L\to\gamma\gamma]$ and $\A[K_L\to\pi^0\to\gamma\gamma]$ coincide.\footnote{At leading order in $SU(3)$ ChPT the $\pi^0$ and $\eta$ contributions cancel (assuming the Gell-Mann--Okubo mass formula), but any realistic $\eta$--$\eta'$ mixing scheme~\cite{GomezDumm:1998gw,DAmbrosio:1994fgc} implies a destructive interference between $\eta$, $\eta'$ and thus $\pi^0$ dominance. The sign of the LD contribution is also discussed in Ref.~\cite{DAmbrosio:2022kvb}.} The sign of $G_8$ cannot be extracted from experiment, but it can be determined by considering hadronic matrix elements of four-quark operators in the $\Delta S=1$ Lagrangian, $\langle\pi^0|\mathcal{H}_W|K_L\rangle$, in a factorization assumption, which leads to $G_8<0$~\cite{Pich:1995qp,Buchalla:2003sj}. This conclusion agrees with large-$N_c$ considerations~\cite{GomezDumm:1998gw,Knecht:1999gb}, and we will therefore adopt 
\beq
\label{norm}
c(0,0)=-3.389(14)\times 10^{-9} \GeV^{-1}
\eeq
as boundary condition. The final representation is then decomposed as 
\beq
\label{rep_final}
c(q_1^{2},q_2^{2})=c^\text{disp}(q_1^{2},q_2^{2})+c^\text{asym}(q_1^{2},q_2^{2}),
\eeq  
where the asymptotic contribution follows as given in Sec.~\ref{sec:matching}
\begin{align}
 c^\text{asym}(q_1^{2},q_2^{2})&=\frac{16\alpha G_F V_{us}^*V_{ud} F_K}{9\pi\sqrt{2}}\big[C_2(\mu)+3C_1(\mu)\big]\bigg\{\frac{m_c^2}{\pi}\int_{s_c}^\infty \diff x\frac{2\pi\log\frac{1+\sigma_c(x)}{1-\sigma_c(x)}}{(x-q_1^2)(x-q_2^2)}\notag\\
 &-\sum_{i=1,2}\frac{q_i^2}{3}\bigg[\int_{s_c}^\infty\diff s\frac{\sigma_c(s)(s+2m_c^2)}{s^2(s-q_i^2)}-\int_{s_u}^\infty \frac{\diff s}{s(s-q_i^2)}\bigg]\bigg\},
\end{align}
with $s_c=4m_c^2$, $s_u=\mathrm{e}^{5/3}m_c^2$, and Wilson coefficients $C_i(\mu)$ including the QCD renormalization group (RG) running from Ref.~\cite{Buchalla:1995vs} to resum the large logarithms $(\alpha_s\log M_W/\mu)^n$ and $\alpha_s(\alpha_s\log M_W/\mu)^n$  to all orders in $\alpha_s$. The scale is set to $\mu^2=|q_1^2+q_2^2|/2$, and the $C_i(\mu)$ are kept constant below $\mu_\text{cut}= 2 \GeV$.
Since the RG corrections to the linear combination $C_2+3C_1$ happen to be large, canceling a significant part of the tree-level result, it is worthwhile to also consider the next-to-leading-order RG corrections from Ref.~\cite{Buchalla:1995vs} to obtain a more realistic estimate of the perturbative uncertainty, despite the dependence on the scheme for $\gamma_5$. We choose the result in the 't Hooft--Veltman scheme~\cite{tHooft:1972tcz} as our central value, and the variation to the leading-order RG and naive dimensional regularization as an estimate of the uncertainty. $c^\text{asym}(0,0)$ then contributes only $2.1\%$ to the form factor normalization, while the correction to the slope is negligible. 

In our final representation, the dispersive part is decomposed according to
\begin{align}
\label{cdisp}
 c^\text{disp}(q_1^{2},q_2^{2})&=c_{vv}(q_1^2,q_2^2)+c_{vs}(q_1^2,q_2^2)+c_{sv}(q_1^2,q_2^2)+c_{ss}(q_1^2,q_2^2)\notag\\
 &+c_{vK^*}(q_1^2,q_2^2)+c_{K^*v}(q_1^2,q_2^2)+c_{sK^*}(q_1^2,q_2^2)+c_{K^*s}(q_1^2,q_2^2),
\end{align}
with
\begin{align}
\label{c_factorized}
 c_{vv}(q_1^2,q_2^2)&=-a_{vv}(q_1^2)a_{vv}(q_2^2),\notag\\
 c_{vs}(q_1^2,q_2^2)&=-a_{vs}(q_1^2)a_{s}(q_2^2),\notag\\
 c_{ss}(q_1^2,q_2^2)&=-a_{ss}(q_1^2,q_2^2),\notag\\
 c_{vK^*}(q_1^2,q_2^2)&=-a_{vv}(q_1^2)\epsilon_{K^*}\frac{ a_{K^{*}}(q_2^2)}{ a_{K^{*}}(0)} 2a_\rho(0)\notag\\
&\times\Re\bigg[\bigg(1+\frac{\beta_{K^{*}}}{3}\bigg)-\frac{ a_{\rho}(q_2^{2})}{ a_\rho(0)}-\frac{\beta_{K^{*}}}{3}\bigg(\frac{1}{3}\frac{ a_{\omega}(q_2^2)}{ a_{\omega}(0)}+\frac{2}{3}\frac{ a_{\phi}(q_2^{2})}{ a_{\phi}(0)}\bigg) \bigg],\notag\\
c_{sK^*}(q_1^2,q_2^2)&=-a_s^{K^*}(q_1^2)\epsilon_{K^*}\frac{ a_{K^{*}}(q_2^2)}{ a_{K^{*}}(0)} 2a_\rho(0)\notag\\
&\times\Re\bigg[\bigg(1+\frac{\beta_{K^{*}}}{3}\bigg)-\frac{ a_{\rho}(q_2^{2})}{ a_\rho(0)}-\frac{\beta_{K^{*}}}{3}\bigg(\frac{1}{3}\frac{ a_{\omega}(q_2^2)}{ a_{\omega}(0)}+\frac{2}{3}\frac{ a_{\phi}(q_2^{2})}{ a_{\phi}(0)}\bigg) \bigg],
\end{align}
$c_{vs}(q_1^2,q_2^2)=c_{sv}(q_2^2,q_1^2)$, $c_{vK^*}(q_1^2,q_2^2)=c_{K^*v}(q_2^2,q_1^2)$, $c_{sK^*}(q_1^2,q_2^2)=c_{K^*s}(q_2^2,q_1^2)$,
and 
\begin{align}
\label{c_detailed}
 a_{vv}(q^2)&=\frac{e N_v}{12\pi^2}\int_{4\mpi^2}^{\sm} \diff x\frac{q_\pi^3(x)[F_\pi^V(x)]^* \Omega_1(x)(1+\alpha_v x)}{x^{1/2}(x-q^2)},\notag\\
 a_{vs}(q^2)&=\frac{e N_s}{12\pi^2}\int_{4\mpi^2}^{\sm} \diff x\frac{q_\pi^3(x)[F_\pi^V(x)]^* \Omega_1(x)(1+\alpha_s x)}{x^{1/2}(x-q^2)},\notag\\
 a_{s}(q^2)&=a_{s}(0)\frac{a_\omega(q^2)}{a_\omega(0)},\qquad a_s^{K^*}(q^2)=\frac{a_{s}(0)}{3}
\bigg[2\frac{a_\phi(q^2)}{a_\phi(0)}+\frac{a_\omega(q^2)}{a_\omega(0)}\bigg],\notag\\
 a_{ss}(q_1^2,q_2^2)&=\frac{a_{ss}(0)}{5}
 \bigg[4\frac{a_\phi(q_1^2)}{a_\phi(0)}\frac{a_\phi(q_2^2)}{a_\phi(0)}+\frac{a_\omega(q_1^2)}{a_\omega(0)}\frac{a_\omega(q_2^2)}{a_\omega(0)}\bigg],
\end{align} 
where 
\begin{align}
 a_{\rho}(q^{2})&= \frac{e}{12\pi^2}\int_{4\mpi^2}^{\sm} \diff x\frac{q_\pi^3(x)[F_\pi^V(x)]^* \Omega_1(x)}{x^{1/2}(x-q^2)}
 =\frac{1}{\pi}\int_{4\mpi^2}^{\sm} \diff x\frac{\rho_\rho(x)}{x-q^2},\notag\\
a_{\omega/\phi/K^*}(q^2)&=\frac{1}{\pi}\int_{\sth}^{\sm} \diff x  \frac{\Im \text{BW}_{\omega/\phi/K^*}(x)}{x-q^2}.
\end{align}

\subsection{Determination of parameters}

The free parameters ($N_{s/v}$, $\alpha_{s/v}$, $a_{s}(0)$, $a_{ss}(0)$, $\epsilon_{K^*}$, $\beta_{K^{*}}$) are determined from the following constraints:
\begin{enumerate}
 \item the fit to $K_L\to\pi^+\pi^-\gamma$, together with the global average for $\alpha_{K^{*}}$ from Ref.~\cite{ParticleDataGroup:2022pth}, determines the three quantities 
$N_{\pi\pi}$, $b_{\pi\pi}$, and $\eta_{K^*}$ as defined in Eq.~\eqref{par_sum},
 \item the total normalization $c^\text{disp}(0,0)+c^\text{asym}(0,0)$ reproduces Eq.~\eqref{norm},
 \item VMD predicts the relative normalizations of $c_{vv}(0,0)$, $c_{vs}(0,0)$, and $c_{ss}(0,0)$ as in Eq.~\eqref{nor_ratio} and the ratio $a_{vv}(0)/a_s(0)$ as in Eq.~\eqref{avv_as_Kstar},
 \item the total slope reproduces Eq.~\eqref{slope}.
\end{enumerate}
To determine the number of independent parameters, we first observe that
\beq
a_\rho(0)N_{\pi\pi}+\bar a_\rho b_{\pi\pi}=\big[a_{vv}(0)\big]^2+a_{s}(0)a_{vs}(0),\qquad \bar a_\rho=\frac{1}{\pi}\int_{4\mpi^2}^{\sm} \diff x\,\rho_\rho(x).
\eeq
Accordingly, 
 the quantities in Eq.~\eqref{par_sum} (with results collected in Table~\ref{tab:fit}), the total normalization~\eqref{norm}, and one of the relative VMD weights given in Eq.~\eqref{nor_ratio} allow one to separate
 the individual terms in 
\beq
c^\text{disp}(0,0)=-\big[a_{vv}(0)\big]^2-2a_{s}(0)a_{vs}(0)-a_{ss}(0,0),
\eeq
and choosing either one of the three relative VMD ratios serves as an estimate of the theoretical uncertainty. 
The product $a_{s}(0)a_{vs}(0)$ is resolved by employing Eq.~\eqref{avv_as_Kstar}.

Next, we observe that 
\beq
\label{singly_virtual}
a_{vv}(q^2)a_{vv}(0)+a_{vs}(q^2)a_{s}(0)=a_\rho(q^2)N_{\pi\pi}+\big(a_\rho(q^2)q^2+\bar a_\rho\big)b_{\pi\pi},
\eeq
which, together with the normalizations $a_{vv}(0)$, $a_{vs}(0)$, $a_{s}(0)$, and $a_{ss}(0,0)$, is sufficient to determine the singly-virtual dependence in the first line of Eq.~\eqref{cdisp}, since $a_s(q^2)$ and $a_{ss}(q_1^2,q_2^2)$ do not involve further free parameters. In contrast, $a_{vv}(q^2)$ and $a_{vs}(q^2)$ are more difficult to disentangle. The definitions of $N_{\pi\pi}$, $b_{\pi\pi}$ in Eq.~\eqref{par_sum} and of $a_{vv}(0)$, $a_{vs}(0)$ in Eq.~\eqref{c_detailed} imply the singular system of equations
\begin{align}
 \begin{pmatrix}
  a_{vv}(0) & a_s(0) & 0 & 0\\
  0& 0& a_{vv}(0)&a_s(0)\\
  a_\rho(0) & 0 & \bar a_\rho & 0\\
  0&a_\rho(0)& 0 & \bar a_\rho
 \end{pmatrix}
\begin{pmatrix}
 N_v\\ N_s\\ N_v\alpha_v \\N_s\alpha_s
\end{pmatrix}
=\begin{pmatrix}
  N_{\pi\pi}\\
  b_{\pi\pi}\\
  a_{vv}(0)\\
  a_{vs}(0)
 \end{pmatrix},
\end{align}
so that not all parameters $N_{s/v}$, $\alpha_{s/v}$ can be determined. For instance, a given solution remains invariant under
\begin{align}
N_v&\to N_v+\eps,& N_s&\to N_s-\eps \frac{a_{vv}(0)}{a_s(0)},\notag\\ 
N_v\alpha_v&\to N_v\alpha_v-\eps\frac{a_\rho(0)}{\bar a_\rho},&
N_s\alpha_s&\to N_s\alpha_s+\eps \frac{a_{vv}(0)a_\rho(0)}{a_s(0)\bar a_\rho}.
\end{align}
Under this transformation, the resulting form for $a_{vv}(q^2)$ changes by
\begin{align}
a_{vv}(q^2)&\to a_{vv}(q^2)+\eps \bigg[a_\rho(q^2)-a_\rho(0)-\frac{a_\rho(0)}{\bar a_\rho}q^2 a_\rho(q^2)\bigg]\notag\\
&=a_{vv}(q^2)-\frac{\eps}{\bar a_\rho}\frac{q^2}{\pi^2}\int_{4\mpi^2}^{\sm} \diff x\int_{4\mpi^2}^{\sm} \diff y\,\rho_\rho(x)\rho_\rho(y)\frac{x-y}{x y(x-q^2)}.
\end{align}
The shift therefore vanishes at $q^2=0$ (by construction), for large $q^2$ (since then the integrand becomes odd under $x\leftrightarrow y$), and in the limit of a narrow resonance (since then the spectral functions force $x-y$ to zero). In practice, we indeed see that the remaining ambiguity in $a_{vv}(q^2)$ and $a_{vs}(q^2)$, which, due to Eq.~\eqref{singly_virtual}, can only affect the doubly-virtual form factor, is completely negligible for reasonably chosen parameter space, so that for all practical purposes the entire functional form of $c^\text{disp}(q_1^2,q_2^2)$ is predicted.

The scale $\Lambda_{\pi\pi}$ in Eq.~\eqref{c_detailed}, beyond which the linear polynomials are continued as constants, is chosen at $\Lambda_{\pi\pi}=\mk$.  With this assignment, our representation saturates the normalization sum rule predicted by the VMD ratios~\eqref{nor_ratio} at a level around 115\%. In practice, to estimate uncertainties we impose the sum-rule normalization exactly and choose one of the three relative weights in Eq.~\eqref{nor_ratio} as predicted from VMD. The average of the three variants is taken as central value, and the spread as a measure of the systematic uncertainty.  Similarly, we take $c_\phi=2c_\omega$ as central value for the $K^*$ piece, see Sec.~\ref{sec:VMD}, but include the variation to $c_\phi=-2c_\omega$ and $c_\phi=0$ in the error estimate.

\section{Prediction for $\boldsymbol{K_L\to\ell^+\ell^-}$}
\label{sec:KL}

\subsection{Reduced amplitude}

 The reduced amplitude $\A_\ell(\mk^2)$ in the prediction for the normalized branching fraction as defined in Eq.~\eqref{R_ell} is typically expressed as
\beq
\label{eq:RD-RA}
\A_\ell(\mk^2) = \frac { 2 i } { \pi^2 \mk^2 } \int \diff^{ 4 } k \frac {\mk^2 k ^ { 2 } - ( q \cdot k ) ^ { 2 } } {  k ^ { 2 }( q-k ) ^ { 2 }\big[ (p-k) ^ { 2 } - m_\ell^{2}\big]} \tilde	c(k^{2},(q-k)^{2}) ,
\eeq
where $q$ is the momentum of  the kaon and  $\tilde c(q_1^{2},q_2^{2})=c(q_1^{2},q_2^{2})/c(0,0)$ is the normalized transition form factor. Therefore, we need to perform the integral for the final representation~\eqref{rep_final} to obtain $\A_\ell(\mk^2)$.  For the isospin part, we can evaluate its contributions following Refs.~\cite{Masjuan:2015cjl, Hoferichter:2021lct}, e.g., for the isovector--isovector piece, we can write it as
\beq
\label{eq:RD-NM} 
\frac {1} { \pi^2 }  \int_{4\mpi^2}^{\sm} \diff x \int_{4\mpi^2}^{\sm} \diff y\, \frac{\tilde \rho(x,y)}{xy} {K(x,y)},\qquad \tilde\rho(x,y)=-\frac{\rho_{vv}(x)\rho_{vv}(y)}{c(0,0)},
\eeq
where $K(x, y)$ is the integration kernel. For the $K^*$ contribution, we need to account for a new topology $K(x, y,z)$ in addition to performing the integral over the pertinent spectral densities.  The integration kernels are spelled out in App.~\ref{app:kernel}, and their reductions and numerical evaluations  are performed using \textit{FeynCalc}~\cite{Shtabovenko:2020gxv,Shtabovenko:2016sxi,Mertig:1990an},  \textit{LoopTools}~\cite{Hahn:1998yk}, and \textit{Package-X}~\cite{Patel:2015tea}. The theoretical uncertainty of the dispersive representation is taken into account as follows: first, the matching threshold $\sqrt{\sm}$ is varied in the range $2.0(2)\GeV$;  fit uncertainties are taken into account by varying the parameters of Table~\ref{tab:fit} in their respective ranges (but the effect turns out to be small); the dependence on VMD assumptions is tested using the different variants for the VMD weights in the normalization as given in  Eq.~\eqref{nor_ratio}, see Sec.~\ref{sec:final}, and for the relative size of $\omega$ and $\phi$ contributions in the $K^*$ piece, see Sec.~\ref{sec:VMD};   lastly,  the transition point $\Lambda_{\pi\pi}$ above which the polynomials~\eqref{polynomials} in the evaluation of $a_{vv}(q^2)$ and $a_{vs}(q^2)$ are kept constant  is varied between $\mk$ and  $1\GeV$. The dispersive uncertainty is then defined as the maximum difference to the central result. 

The asymptotic contributions from the three-point and  two-point functions read
\begin{align}
	\label{Aasym}
	\A_\ell^{\text{asym}, I}(\mk^2)&=\frac{16\alpha G_F V_{us}^*V_{ud} F_Km_c^2}{9\pi^2\sqrt{2}\,c(0,0)}\big[C_2(\mu)+3C_1(\mu)\big] \int_{s_c}^\infty \diff x\frac{2\pi\log\frac{1+\sigma_c(x)}{1-\sigma_c(x)}}{x^2}K(x,x),\notag\\
	\A_\ell^{\text{asym}, T}(\mk^2)&=\frac{32\alpha G_F V_{us}^*V_{ud} F_K}{9\pi^2\sqrt{2}\,c(0,0)}\big[C_2(\mu)+3C_1(\mu)\big]\int_{s_c}^\infty \diff x\frac{\Im T_q(x)}{x^2}K(x)-(c\leftrightarrow u),
\end{align}
where the definition of $K(x)$ is also collected in App.~\ref{app:kernel}. In practice,  we estimate the asymptotic contributions by means of the approximate expressions given in App.~\ref{app:kernel} to be able to include the running of the Wilson coefficients in a straightforward manner. We have crosschecked that the approximate formula~\eqref{Aq2} indeed reproduces the results of the exact calculations~\eqref{Aasym} at tree level, up to tiny corrections that are entirely negligible compared to the dominant perturbative uncertainties.\footnote{To go beyond this approximation, one would need to express the momentum dependence of the Wilson coefficients in a form amenable to a direct evaluation of the loop integral. This could be achieved, for instance, by approximating the RG solution with Pad\'e approximants, but in view of the size of the corrections observed for constant Wilson coefficients such effects are not relevant at the present level of precision.}  Therefore, we vary the scheme  and the order of  perturbation theory to estimate the asymptotic uncertainty: it is defined as the maximum deviation of calculations either in the naive dimensional regularization scheme or at the leading-log approximation from the central result evaluated in the 't Hooft--Veltman scheme. This turns out to be a conservative choice as it amounts to a $3\%$ violation of the form factor normalization, well beyond the experimental precision~\eqref{norm}.

With these procedures for uncertainty estimates, we find the following imaginary parts of the amplitude
\beq
\label{ImA_final}
\Im\A_\mu(\mk^2)=-5.20(0), \qquad \Im\A_e(\mk^2)=-21.59(1).
\eeq
Beyond the $\gamma\gamma$ imaginary parts,
\beq
\label{ImA_gg}
\text{Im}_{\gamma\gamma}\A_\mu(\mk^2)=-5.21, \qquad \text{Im}_{\gamma\gamma}\A_e(\mk^2)=-21.62, 
\eeq
our analysis also takes the additional $\pi\pi\gamma$ and $3\pi\gamma$ intermediate states into account via the dispersively reconstructed spectral functions, but the comparison of Eqs.~\eqref{ImA_final} and~\eqref{ImA_gg} reaffirms that the imaginary parts are entirely dominated by the $\gamma \gamma$ cut~\cite{Martin:1970ai}. For the total LD contribution to the real parts,  we obtain
\begin{align}
	\label{ReA_final}
	\Re\A_\mu^\text{LD}(\mk^2)&=-0.50_{\text{disp}}+0.34_{\text{asym}}=-0.16(21)_{\text{disp}}(27)_{\text{asym}}(17)_{\text{exp}}[38]_\text{total},\notag\\
	 \Re\A^\text{LD}_e(\mk^2)&=31.99_{\text{disp}}-0.31_{\text{asym}}=31.68(59)_{\text{disp}}(73)_{\text{asym}}(27)_{\text{exp}}[98]_\text{total},
\end{align}
where the errors refer to the uncertainties in the dispersive and asymptotic part of the representation, respectively, as well as the experimental uncertainty in the slope~\eqref{slope}. Matching to Eq.~\eqref{ReA}, this translates to the low-energy constants
\beq
\label{LEC}
\chi_\mu(\mu)=4.96(38), \qquad \chi_e(\mu)=8.0(1.0), 
\eeq
evaluated at $\mu=0.77\GeV$. Accordingly, extracting $\chi(\mu)$  at one-loop order is not sufficient to obtain the expected lepton-flavor-universal result~\cite{DAmbrosio:1997eof,Isidori:2003ts,Crivellin:2016vjc}, reflecting the impact of higher chiral orders~\cite{Vasko:2011pi,Husek:2014tna}. A similar effect has been observed for $\eta,\eta'\to\ell^+\ell^-$~\cite{Masjuan:2015cjl}, and the comparison to $\chi(\mu)=2.69(10)$ from $\pi^0\to e^+e^-$~\cite{Hoferichter:2021lct} highlights further that one-loop ChPT is not sufficient for an accurate phenomenology of pseudoscalar dilepton decays. Adding the SD contribution~\eqref{chi_SD_SM}, the complete real parts are\footnote{If the sign of $c(0,0)$ in Eq.~\eqref{norm} were positive, the LD contributions would change to 
\begin{align*}
	\label{ReA_final_flipped}
	\Re\A_\mu^\text{LD}(\mk^2)&=-0.32_{\text{disp}}-0.34_{\text{asym}}=-0.66(20)_{\text{disp}}(27)_{\text{asym}}(17)_{\text{exp}}[38]_\text{total},\notag\\
	 \Re\A^\text{LD}_e(\mk^2)&=30.98_{\text{disp}}+0.31_{\text{asym}}=31.29(59)_{\text{disp}}(73)_{\text{asym}}(27)_{\text{exp}}[98]_\text{total},
\end{align*}
and the total real parts would become
\begin{equation*}
  \Re\A_\mu^\text{LD+SD}(\mk^2)=1.14(38),\qquad \Re\A_e^\text{LD+SD}(\mk^2)=33.1(1.0).
\end{equation*}
}   
\beq
\label{ReA_full_final}
\Re\A_\mu^\text{LD+SD}(\mk^2)=-1.96(39),\qquad \Re\A_e^\text{LD+SD}(\mk^2)=29.9(1.0),
\eeq
leading to the branching fractions
\beq
\label{Br_final}
\Br[K_L\to\mu^+\mu^-]=7.44^{+0.41}_{-0.34}\times 10^{-9},\qquad\Br[K_L\to e^+e^-]=8.46(37)\times 10^{-12}.
\eeq 
The final result for $\Re\A_\mu(\mk^2)$ can then be compared to experiment, Eq.~\eqref{RLexp},
\beq
\label{ReAexp}
\Re\A_\mu^\text{exp}(\mk^2)=\pm 1.16(24),
\eeq
where we used Eq.~\eqref{ImA_final} for $\Im \A_\mu$. This leaves room for a BSM component
\beq
\label{ReA_BSM}
\Re\A_\mu^\text{BSM}(\mk^2)=\begin{cases}
3.12(46)\\
0.80(46)
\end{cases},
\eeq
depending on the sign in Eq.~\eqref{ReAexp}. Choosing the negative value, our prediction thus displays a mild tension at the level of $1.7\sigma$ with experiment. For the electron mode, the experimental value reads
\beq
\label{ReAexpe}
\Re\A_e^\text{exp}(\mk^2)=\pm
30.5(13.0),
\eeq
in agreement with Eq.~\eqref{ReA_full_final}, but
with errors too large to derive meaningful BSM constraints.

\subsection{Comparison to previous work}

\begin{figure}[t]
	\centering
	\includegraphics[width=.8\linewidth]{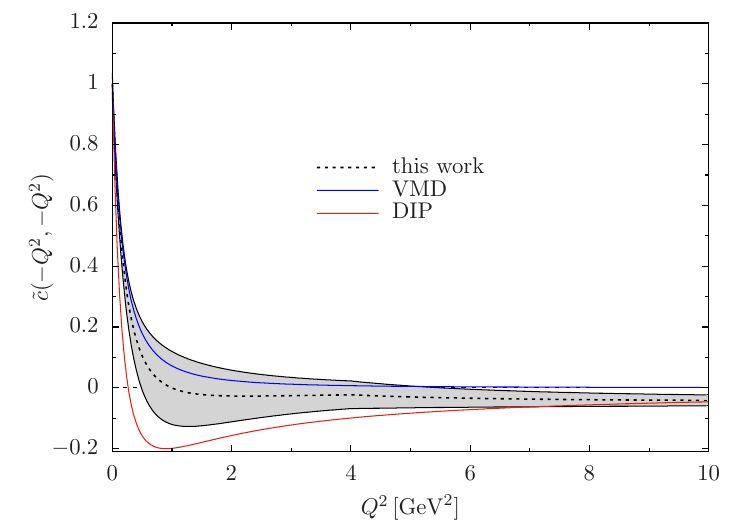}
	\caption{The normalized diagonal form factor $\tilde c(-Q^2,-Q^2)$ (black dashed line with uncertainty band),
		in comparison to the VMD model and the DIP parameterization~\cite{DAmbrosio:1997eof}. }
	\label{fig:FF_diag}
\end{figure} 

It is instructive to compare our results for the LD contribution to previous evaluations, see Fig.~\ref{fig:FF_diag} and Table~\ref{tab:comparison}.\footnote{For the BMS parameterization such a comparison is not possible, since the model only applies to the singly-virtual form factor. Naive factorization would produce terms with two $K^*$ propagators, which cannot occur in the presence of a single electroweak vertex.} The simplest model, a VMD form factor with mass scale fit to reproduce the slope~\eqref{slope}, gives
\beq
	\Re\A_\mu^{\text{LD, VMD}}(\mk^2)=-0.68(11),\qquad \Re\A_e^{\text{LD, VMD}}(\mk^2)=31.36(26),
\eeq
where the uncertainty is propagated from $b_{K_L}$. The corresponding diagonal form factor is compared to our result in Fig.~\ref{fig:FF_diag}, which shows that the main differences to the VMD model concern a faster decrease for small $Q^2$ and a negative asymptotic tail. While the dispersive part in Eq.~\eqref{ReA_final} comes out similar to the VMD result, the asymptotic contribution entails a positive shift that brings the net result closer to the experimental value.  Ultimately, this enhancement arises from logarithmic corrections that lead to an asymptotic behavior proportional to $\log Q^2/Q^2$, a feature that is absent in both the VMD and DIP models, with asymptotic limits proportional to $1/Q^4$ and $1/Q^2$, respectively.  

For the DIP parameterization, with $\alpha_{\text{DIP}}=-1.69(8)$ from the global determination~\cite{ParticleDataGroup:2022pth} and $\beta_\text{DIP}=-1-2\alpha_{\text{DIP}}$ from the asymptotic constraint,  
one finds
\beq
\Re\A_\mu^{\text{LD, DIP}}(\mk^2)=0.57(28), \qquad \Re\A_e^{\text{LD, DIP}}(\mk^2)=32.46(71),
\eeq
propagating the uncertainties from $\alpha_\text{DIP}$. The comparison to our result is again illustrated in Fig.~\ref{fig:FF_diag}, which shows that the higher central value of $\Re\A_\mu(\mk^2)$ is driven by a more pronounced minimum of $\tilde c(-Q^2,-Q^2)$ around $Q^2=1\GeV^2$, more than compensating for the missing $\log Q^2$ in the asymptotic behavior and thereby even reversing the sign of the resulting value for $\Re\A_\mu(\mk^2)$. The corresponding low-energy constant $\chi_\mu^{\text{DIP}}(\mu)=5.70(28)$ comes out close to $\chi_\mu^\text{\cite{Isidori:2003ts}}(\mu)=5.83(1.0)_{\text{th}}(0.15)_{\text{exp}}$. This central value was  obtained in Ref.~\cite{Isidori:2003ts} from a DIP parameterization with $\alpha_\text{DIP}=-1.611(44)$  and an expansion in $\mk$ and $m_\mu$.\footnote{The value for $\alpha_\text{DIP}$ differs from the current PDG average because at the time only a preliminary value $\alpha_\text{DIP}=-1.63(5)$ from $K_L\to e^+e^-\gamma$ was available~\cite{Corcoran:2003njf},  which later changed to $\alpha_\text{DIP}=-1.73(5)$~\cite{KTeV:2007ksh}.} Using the same value for $\alpha_\text{DIP}$, but keeping the full loop integral, we find $\chi_\mu(\mu)=5.42(15)$ for this DIP contribution. The uncertainty is dominated by an estimate of high-energy contributions by varying a phenomenological ansatz within the range allowed by the partonic expression~\eqref{asym}.\footnote{We believe that the sign of the form factor in Fig.~3 of Ref.~\cite{Isidori:2003ts} should be reversed, to be consistent with the sign convention $c(0,0)<0$. However, since the comparison to the partonic calculation is only used for an error estimate, the conclusions remain essentially unchanged.}  
In this work, we improved the matching to asymptotic constraints by separating the low- and high-energy parts of the loop functions via dispersion integrals, to the effect that the main uncertainty now arises from the RG corrections to the Wilson coefficients. However, the range for the high-energy contributions considered in Ref.~\cite{Isidori:2003ts} does suggest a correction to the DIP solution in the direction of the VMD form factor, which is consistent with our findings and would move the value of the low-energy constant closer to Eq.~\eqref{LEC}.  

\begin{table*}[t]
	\centering
	\renewcommand{\arraystretch}{1.3}
	\small
	\begin{tabular}{l r r r r}
		\toprule
		& $\Re\A_\mu^\text{LD}(\mk^2)$ & $\Re\A_e^\text{LD}(\mk^2)$ & $\chi_\mu(\mu)$ & $\chi_e(\mu)$\\
		\midrule
		 VMD & $-0.68(11)$ & $31.36(26)$ & $4.44(11)$ & $7.6(3)$\\
		 DIP & $0.57(28)$ & $32.46(71)$ & $5.70(28)$ & $8.7(7)$\\
		 This work & $-0.16(38)$ & $31.68(98)$ & $4.96(38)$ & $8.0(1.0)$\\
		\bottomrule
	\renewcommand{\arraystretch}{1.0}
	\end{tabular}
	\caption{Comparison of the LD amplitudes and LECs to VMD and DIP parameterizations, see main text for details.}
	\label{tab:comparison}
\end{table*}

\section{Consequences for physics beyond the Standard Model}
\label{sec:BSM}

\subsection{SMEFT}

BSM constraints are most easily expressed in terms of the Wilson coefficients in
\beq
{\mathcal L}^\text{BSM}=C_A^\ell \bar s \gamma^\mu \gamma_5 d\,\bar\ell\gamma_\mu\gamma_5\ell+C_P^\ell\bar si\gamma_5 d\,\bar\ell i\gamma_5\ell+\text{h.c.},
\eeq
leading to
\begin{align}
\Re \A_\ell^\text{BSM}&=-\frac{4m_\ell F_K}{\mathcal{N}}\bigg(\Re C_A^\ell+\frac{\mk^2}{2m_\ell(m_s+m_d)}\Re C_P^\ell\bigg)\notag\\
&=1.124(5)\times 10^{11}\GeV^2\bigg(\Re C_A^\ell+\frac{\mk^2}{2m_\ell(m_s+m_d)}\Re C_P^\ell\bigg),
\end{align}
where the normalization
\beq
\mathcal{N}=-\frac{\alpha}{\pi}\frac{m_\ell}{\mk}\sqrt{\frac{16\pi\Gamma[K_L\to\gamma\gamma]}{\mk}}
\eeq
takes into account the relative sign compared to the LD contribution in the same conventions as in Sec.~\ref{sec:final} and App.~\ref{app:SD} (similar expressions are available in the literature, see, e.g., Refs.~\cite{Mescia:2006jd,Chobanova:2017rkj}). The axial-vector coefficient maps onto the Gilman--Wise basis~\cite{Gilman:1979bc,Gilman:1979ud,Buras:1998raa}
\beq
{\mathcal L}=-\frac{G_F}{\sqrt{2}}V_{us}^*V_{ud} C_{7A}^\ell\,\bar s\gamma^\mu(1-\gamma_5)d\,\bar\ell\gamma_\mu\gamma_5\ell+\text{h.c.}
\eeq
according to $C_A^\ell=\frac{G_F}{\sqrt{2}} V_{us}^*V_{ud} C_{7A}^\ell$, and the tree-level matching to SMEFT coefficients~\cite{Grzadkowski:2010es,Buchmuller:1985jz} reads
\beq
C_A^\ell=\frac{1}{4}\Big(C^{(1)}_{\ell q}+C^{(3)}_{\ell q}+C_{ed}-C_{\ell d}-C_{qe}\Big),\qquad 
C_P^\ell=-\frac{1}{4} C_{\ell e d q},
\eeq
with flavor indices $prst=\ell\ell21$.

\subsection[Modified $Z$ couplings]{Modified $\boldsymbol{Z}$ couplings}

A special case is given by the test of modified $Z$ couplings, defined by the effective Lagrangian~\cite{Buras:1998ed}
\beq
{\mathcal L}=\frac{G_F}{\sqrt{2}}\frac{e}{2\pi^2}M_Z^2\frac{\cos\theta_W}{\sin\theta_W} Z_{ds}\,\bar s\gamma^\mu(1-\gamma_5)d\, Z_\mu,
\eeq
which maps to
\beq
C_A^\ell=\frac{\alpha(M_Z)G_F}{2\pi\sqrt{2}\,\sin^2\theta_W} Z_{ds}. 
\eeq
The SM contribution to $Z_{ds}$, corresponding to the last two diagrams in Fig.~\ref{fig:diagrams_SD},  gives $Z_{ds}=\lambda_t C_0(x_t)$~\cite{Buchalla:1990qz}, and, together with the $W$ box, determines the SD part of $\chi(\mu)$ in Eq.~\eqref{chi_SD_SM}. Choosing the sign in Eq.~\eqref{ReA_BSM} in line with the SM, we obtain the constraint
\beq
\label{ReZds}
\Re Z_{ds}^\text{BSM}=1.6(5)_\text{exp}(8)_\text{SM}\times 10^{-4},\qquad |\Re Z_{ds}^\text{BSM}|<2.8\times 10^{-4} \quad \text{at } 90\%\text{ C.L.}
\eeq
Limits on $\Re Z_{ds}$ can also be extracted from $K^+\to\pi^+\nu\bar\nu$~\cite{NA62:2021zjw} 
\beq
\Br[K^+\to\pi^+\nu\bar\nu]\big|_\text{NA62}=\big(10.6^{+4.0}_{-3.4}\pm 0.9\big)\times 10^{-11},
\eeq
which is sensitive to both the real and imaginary part of $Z_{ds}$~\cite{Buras:1998ed,Buras:2015qea,Brod:2021hsj}
\begin{align}
\Br[K^+\to\pi^+\nu\bar\nu]&=\kappa_+\big(1+\Delta_\text{EM}\big)\bigg[\bigg(\frac{\Im\lambda_t X_t+\Im Z_{ds}^\text{BSM}}{\lambda^5}\bigg)^2\notag\\
&+\bigg(\frac{\Re\lambda_c}{\lambda}\big(P_c+\delta P_{c,u}\big)+\frac{\Re\lambda_t X_t+\Re Z_{ds}^\text{BSM}}{\lambda^5}\bigg)^2
\bigg].
\end{align}
Evaluating the SM contribution with the input quantities summarized in Ref.~\cite{Brod:2021hsj} (relying on Refs.~\cite{Misiak:1999yg,Buchalla:1998ba,Isidori:2005xm,Buras:2006gb,Mescia:2007kn,Brod:2008ss,Brod:2010hi}), but with the CKM parameters as collected in App.~\ref{app:SD}, we find $\Br[K^+\to\pi^+\nu\bar\nu]|_{\text{SM}}=8.2(5)\times 10^{-11}$, in good agreement with Ref.~\cite{CERN2023}. Assuming $\Im Z_{ds}^\text{BSM}=0$ and choosing again the sign that ensures agreement with the SM, we obtain
\beq
\Re Z_{ds}^\text{BSM}=-1.1(1.5)_\text{exp}(0.2)_\text{SM}\times 10^{-4},\qquad |\Re Z_{ds}^\text{BSM}|<3.1\times 10^{-4} \quad \text{at } 90\%\text{ C.L.},
\eeq
leading to a limit almost identical to Eq.~\eqref{ReZds}. This observation nicely illustrates the complementarity of $K_L\to\mu^+\mu^-$ and $K^+\to\pi^+\nu\bar \nu$, since a combined analysis of the two processes can be used to disentangle BSM contributions to the real and imaginary part of $Z_{ds}$ at a similar level of precision. In view of the projected experimental advances for $K^+\to\pi^+\nu\bar\nu$ at NA62 and HIKE, this motivates commensurate efforts to improve the SM prediction for $K_L\to\mu^+\mu^-$ as well.

\section{Summary and outlook}
\label{sec:summary}

In this work, we analyzed the $K_L\to\gamma^*\gamma^*$ form factor in dispersion theory. To predict its low-energy part, we made use of data for $K_L\to\gamma\gamma$ and $K_L\to\ell^+\ell^-\gamma$, $K_L\to\ell_1^+\ell_1^-\ell^+_2\ell^-_2$ to determine normalization and slope, respectively, but, crucially, also for $K_L\to\pi^+\pi^-\gamma$, to constrain the $2\pi$ cuts in the spectral function for an isovector photon. Together with minimal narrow-width parameterizations for the isoscalar spectral functions as well as the $K^*\simeq K\pi$ contributions, we found that the resulting form factor, albeit mainly constrained by singly-virtual data, leaves remarkably little flexibility for doubly-virtual kinematics either. The representation for the full form factor was then supplemented by an asymptotic contribution, which derives from a dispersive formulation of the partonic calculation of $K_L\to\gamma^*\gamma^*$, to be able to separate low- and high-energy contribution by a suitable cutoff in the integrals. We observed that besides uncertainties in the dispersive representation and from the experimental value of the slope parameter, another important source of uncertainty arises from the perturbative running of the Wilson coefficients, which we studied including the resummation of subleading logarithms. 

The central results are given in Eq.~\eqref{ReA_final} for the real part of the long-distance amplitudes $\Re\A_\ell(\mk^2)$, and in Eq.~\eqref{Br_final} for the corresponding branching fractions. The sum of the long-distance amplitude for $K_L\to\mu^+\mu^-$ and the short-distance contribution in the SM agrees with experiment at the level of $1.7\sigma$, and the improved uncertainty in $\Re\A_\mu(\mk^2)$ allows us to formulate more stringent limits on physics beyond the SM. The uncertainty in the prediction of the long-distance contribution could be further improved if better data for the $K_L\to\pi^+\pi^-\gamma$ and $K_L\to\ell^+\ell^-\gamma$ spectra were available, this would allow one to reduce the dispersive uncertainty and the precision of the form-factor slope. In this context, we emphasize that it is unfortunate that the original data for the respective spectra 
from Refs.~\cite{KTeV:2007ksh,KTeV:2001sfq,E731:1992nnl,KTeV:2006diq} are not accessible, as, especially for the dilepton data, we are thus forced to rely on the quoted parameters for fits using a specific model for the form factor. While we tried to minimize the impact of the corresponding assumptions by constructing our dispersive parameterization in a way consistent with these fits, they could have been avoided altogether if the data for the spectra themselves had been preserved.

The asymptotic uncertainty could potentially be improved using the interplay with lattice QCD~\cite{Chao:2023cxp,Christ:2022rho,Zhao:2022pbs,Christ:2020bzb,Hoid:2023has}, which could also help unambiguously determine the relative sign of long- and short-distance contributions. Further, it would be very helpful if lattice QCD could provide information on the relative weights between normalizations of different isospin currents, as this would allow one to test the vector-meson-dominance assumptions currently needed for the prediction of the doubly-virtual form factor.   
Already with the improvements implemented in this work, the final uncertainty in Eq.~\eqref{ReA_full_final} is getting close to experiment~\eqref{ReAexp}, so that with future advances in theory it should become possible to match the current experimental precision, especially, if improved input for the $K_L\to\pi^+\pi^-\gamma$ and $K_L\to\ell^+\ell^-\gamma$ spectra became available. As we argued in Sec.~\ref{sec:BSM}, the corresponding constraints on physics beyond the Standard Model are complementary to the ones obtained from $K\to\pi\nu\bar\nu$, motivating further efforts to improve the SM prediction for $K_L\to\mu^+\mu^-$.

\acknowledgments
We thank Michael Akashi-Ronquest and John Belz for correspondence on Refs.~\cite{KTeV:2000avq,KTeV:2006diq}, and Gino Isidori for discussions on Ref.~\cite{Isidori:2003ts}. We further thank Joachim Brod, Giancarlo D'Ambrosio, Martin Gorbahn, and Marc Knecht for discussions during the meeting Kaons@CERN 2023.  
Financial support by the SNSF (Project No.\ PCEFP2\_181117) and by the Ram\'on y Cajal program (RYC2019-027605-I) of the Spanish MINECO is gratefully acknowledged. 

\appendix

\section{Short-distance contribution in the Standard Model}
\label{app:SD}

Following Refs.~\cite{Buchalla:1993wq,Gorbahn:2006bm}, the SD contribution in the SM can be expressed as
\beq
\label{K_L_SD}
 \Br[K_L\to\mu^+\mu^-]\big|_\text{SD}=\kappa_\mu\bigg[\frac{\Re\lambda_t}{\lambda^5}Y(x_t)+\frac{\Re \lambda_c}{\lambda}P_c\bigg]^2,
\eeq
where $\lambda_i=V_{is}^*V_{id}$ collects the CKM matrix elements, $\lambda=|V_{us}|$ denotes the Wolfenstein parameter, $Y(x_t)$ is a loop function depending on the top quark $\overline{\text{MS}}$ mass via $x_t=[m_t(\mu_t)]^2/M_W^2$ and the top quark matching scale $\mu_t$, and $P_c$ gives the contribution from the charm loop. The dependence on $\lambda$ is far less pronounced than Eq.~\eqref{K_L_SD} suggests, since the normalization
\beq
\label{kappa_mu}
\kappa_\mu=\frac{[\alpha(M_Z)]^2\Br[K^+\to\mu^+\nu_\mu]}{\pi^2\sin^4\theta_W}\frac{\tau(K_L)}{\tau(K^+)}\lambda^8=1.999(9)\times 10^{-9}\bigg(\frac{\lambda}{0.225}\bigg)^8
\eeq
together with the scaling $P_c\propto \lambda^{-4}$ cancels all but the remaining power $1/\lambda^2$ in $\Br[K_L\to\mu^+\mu^-]$. For the numerical evaluation of Eq.~\eqref{kappa_mu} we use the global fit from Ref.~\cite{Cirigliano:2022yyo},
\beq
\Br[K^+\to\mu^+\nu_\mu]=63.58(11)\%,\qquad \tau(K^+)=12.384(15)\,\text{ns},\qquad  \tau(K_L)=51.16(21)\,\text{ns},
\eeq
as well as $\alpha(M_Z)\equiv\alpha_{\overline{\text{MS}}}(M_Z)=1/127.9$ and $\sin^2\theta_W\equiv \sin^2\hat\theta_W^{\overline{\text{MS}}}=0.23122(4)$~\cite{ParticleDataGroup:2022pth}.\footnote{For the LD part we always use $\alpha\equiv\alpha(0)=1/137.036$.} For the CKM matrix elements we use~\cite{ParticleDataGroup:2022pth,CKM2023}
\begin{align} 
\Re\lambda_c&=-\lambda\bigg(1-\frac{\lambda^2}{2}\bigg)=-0.2193(7),\qquad \lambda=0.2250(8),\notag\\
\Re\lambda_t&=-A^2\lambda^5\bigg(1-\frac{\lambda^2}{2}\bigg)(1-\bar\rho)=-3.21(11)\times 10^{-4},\notag\\
A&=0.824(11),\qquad \bar\rho=0.158(10),\notag\\
\Im\lambda_t&=A^2\bar\eta \lambda^5\bigg(1+\frac{\lambda^2}{2}\bigg)=1.41(6)\times 10^{-4},\qquad \bar\eta=0.351(9),
\end{align}
where we have taken the mean of the numbers using CKMFitter~\cite{Hocker:2001xe,Charles:2004jd} and UTfit~\cite{UTfit:2005ras,UTfit:2007eik} methodology and assigned the respective larger error. 

The loop function $Y(x_t)$ has recently been updated in Ref.~\cite{Brod:2022khx}, and we use the resulting value $Y(x_t)=0.931(5)$. For $P_c$ we use the approximate formula given in Ref.~\cite{Gorbahn:2006bm} to update its value to $m_c(m_c)=1.278(13)\GeV$~\cite{FlavourLatticeAveragingGroupFLAG:2021npn,EuropeanTwistedMass:2014osg,Alexandrou:2014sha,Chakraborty:2014aca,FermilabLattice:2018est,Hatton:2020qhk} and $\alpha_s(M_Z)=0.1184(8)$~\cite{FlavourLatticeAveragingGroupFLAG:2021npn,Chakraborty:2014aca,Maltman:2008bx,PACS-CS:2009zxm,McNeile:2010ji,Bruno:2017gxd,Bazavov:2019qoo,Cali:2020hrj,Ayala:2020odx}, which results in
\beq
P_c=0.111(10)\bigg(\frac{0.225}{\lambda}\bigg)^4. 
\eeq
The error therein is dominated by the residual scale ambiguities at this order, to which we have added linearly the small remaining uncertainty $\Delta P_c=0.002$ due to $m_c$, see Ref.~\cite{Gorbahn:2006bm}. Taking everything together, we obtain
\beq
\Br[K_L\to\mu^+\mu^-]\big|_\text{SD}=0.784(55)\times 10^{-9},
\eeq
which translates to the SD contribution
\beq
\label{chi_SD_1}
\chi_\text{SD}^\text{SM}=-1.804(64). 
\eeq
The uncertainty is completely dominated by the CKM matrix elements, with the largest contribution from the Wolfenstein parameter $A$. 

An equivalent representation of $\chi_\text{SD}^\text{SM}$ was derived in Ref.~\cite{Isidori:2003ts} starting from 
\beq
C_\text{SD}^\text{SM}=-\frac{\sqrt{2}\,G_F\alpha(M_Z)m_\mu F_K}{\pi\sin^2\theta_W}\Big(\Re\lambda_t Y(x_t)+\lambda^4\Re\lambda_c  P_c\Big),
\eeq
which amounts to
\begin{align}
\label{chi_SD_2}
\chi_\text{SD}^\text{SM}&=-\tilde \kappa\Big(\Re\lambda_t Y(x_t)+\lambda^4 \Re\lambda_c P_c\Big)=-1.802(59),\notag\\
|\tilde\kappa|&=\sqrt{\frac{\mk}{16\pi\Gamma[K_L\to\gamma\gamma]}}\frac{\sqrt{2}\,G_F\mk F_K\alpha(M_Z)}{\sin^2\theta_W \alpha(0)}=4988.4(23.6),
\end{align}
where we have used $G_F=1.1663788(6)\times 10^{-5}\GeV^{-2}$~\cite{ParticleDataGroup:2022pth,MuLan:2012sih} and $F_K=110.58(24)\MeV$ as additional inputs.\footnote{This value uses $F_K/F_\pi=1.1978(22)$ in the isospin limit~\cite{Dowdall:2013rya,Bazavov:2017lyh,Miller:2020xhy,ExtendedTwistedMass:2021qui,Cirigliano:2022yyo} together with $F_\pi=92.32(10)\MeV$~\cite{ParticleDataGroup:2022pth}, which should be close to the input required for the neutral-kaon decay. The formulation in Eq.~\eqref{kappa_mu} evades this subtlety by instead expressing the kaon decay constant in terms of $\Br[K^+\to\mu^+\nu_\mu]$, but neither variant accounts for isospin-breaking effects in a consistent manner. However, the agreement between Eqs.~\eqref{chi_SD_1} and \eqref{chi_SD_2} shows that the resulting error is negligible compared to the dominant CKM uncertainties.} The resulting value of $\chi_\text{SD}^\text{SM}$ is in good agreement with Eq.~\eqref{chi_SD_1}, motivating the final estimate quoted in Eq.~\eqref{chi_SD_SM}. 
In both cases we have introduced a relative sign to account for the negative normalization of the $K_L\to\gamma\gamma$ amplitude, see Sec.~\ref{sec:final}.

\section{Dispersive representation of loop functions}
\label{app:loop}

\subsection{Two-point function}

The two-point function for quark flavor $q=u,c$ can be represented in the form
\beq
T_q(q^2)=2\int_0^1  \diff x \,x(1-x)\log\bigg[1-x(1-x)\frac{q^2}{m_q^2}\bigg],
\eeq
and recast into the dispersive representation
\beq
\label{T_disp}
T_q(q^2)=\frac{q^2}{\pi}\int_{4m_q^2}^\infty \diff s \frac{\Im T_q(s)}{s(s-q^2)},
\eeq
with
\beq
\Im T_q(s)=-\frac{\pi}{3}\sigma_q(s)\bigg(1+\frac{2m_q^2}{s}\bigg). 
\eeq
In the matching to short-distance constraints, we use a variant of Eq.~\eqref{T_disp} in which $m_u\to 0$ and lower cutoffs $s_q$ applied in the dispersion integrals, to separate the parts already accounted for via the low-energy contributions. This leads to 
\begin{align}
\label{TcTu}
T_c^\text{asym}(q^2)-T_u^\text{asym}(q^2)&=-\frac{q^2}{3}\bigg[\int_{s_c}^\infty\diff s\frac{\sigma_c(s)(s+2m_c^2)}{s^2(s-q^2)}-\int_{s_u}^\infty \frac{\diff s}{s(s-q^2)}\bigg]\notag\\
&=-\frac{1}{3}\bigg[\frac{5}{3}\big(1-\sigma_c(s_c)\big)+\frac{4m_c^2}{q^2}\bigg(1-\sigma_c(s_c)\Big(1+\frac{q^2}{3s_c}\Big)\bigg)\notag\\
&\quad+\frac{\sigma_c(q^2)(q^2+2m_c^2)}{q^2}\log\frac{\frac{\sigma_c(q^2)+\sigma_c(s_c)}{\sigma_c(q^2)-\sigma_c(s_c)}}{\frac{\sigma_c(q^2)+1}{\sigma_c(q^2)-1}}
+\log\frac{s_u-q^2}{s_u}\bigg]\notag\\
&=-\frac{1}{3}\bigg[\frac{5}{3}\big(1-\sigma_c(s_c)\big)-\frac{4m_c^2\sigma_c(s_c)}{3s_c}+\log\frac{1+\sigma_c(s_c)}{1-\sigma_c(s_c)}-\log\frac{s_u}{m_c^2}\bigg]\notag\\
&\quad+\Order\Big(\frac{1}{q^2}\Big),
\end{align}
written in a form that applies for space-like $q^2$ and expanded for asymptotic values in the last step. To ensure that the entire contribution drops as $1/q^2$, $s_u$ and $s_c$ therefore cannot be varied independently. In particular, one has $s_u>s_c$, e.g., for $s_c=4m_c^2$ one needs to set $s_u=\mathrm{e}^{5/3}m_c^2$.

\subsection{Three-point function}

In the literature, the three-point function is represented in the form~\cite{Simma:1990nr,Herrlich:1991bq}
\begin{align}
\label{Iq_lit}
 	I_q(r_1,r_2,r_3)&=\int_0^1 \diff x\int_0^{1-x}\diff y\bigg\{\frac{2-x-y+(r_1 x+ r_2 y)(1-x-y)^2+r_3xy (2-x-y)}{1-(r_1 x+r_2 y)(1-x-y)-r_3xy }\notag\\
	&\quad-(2-3x-3y)\log\big[1-(r_1 x+r_2 y)(1-x-y)-r_3xy \big]\bigg\}-\frac{2}{3},
\end{align}
where $r_1=q_1^2/m_q^2$, $r_2=q_2^2/m_q^2$, $r_3=\mk^2/m_q^2$. Starting from this parameterization, we calculated the discontinuity in $r_1$, which takes a very simple form
\beq
\label{discIq}
\text{disc}_{r_1} I_q(r_1,r_2,r_3)=-\frac{4\pi i\theta(r_1-4)}{\lambda^{1/2}_{123}}\log\frac{r_1-r_2-r_3-\sigma_1\lambda_{123}^{1/2}}{r_1-r_2-r_3+\sigma_1\lambda_{123}^{1/2}},
\eeq
where $\lambda_{123}=\lambda(r_1,r_2,r_3)$ and $\sigma_1=\sqrt{1-4/r_1}$. In particular, this expression coincides up to a factor $-2$ with the corresponding discontinuity of the scalar loop function 
\beq
\label{C0}
C_0(r_1,r_2,r_3)=-\int_0^1\diff x\int_0^{1-x}\diff y\frac{1}{1-(r_1 x+r_2 y)(1-x-y)-r_3xy},
\eeq
so that the difference to Eq.~\eqref{Iq_lit} can only arise from a subtraction polynomial. Indeed, we find that 
\beq
\label{Iq_simp}
I_q(r_1,r_2,r_3)=-2C_0(r_1,r_2,r_3)-1
\eeq
holds in general, and will continue to work with this simplified expression. The loop function fulfills the limits
\beq
 I_q(0,0,r_3)=\frac{4}{r_3}\arctan^2\frac{\sqrt{r_3}}{\sqrt{4-r_3}}-1=\frac{r_3}{12}+\Order\big(r_3^2\big)
\eeq
and
\beq
I_q(r_1,r_2,r_3)\big|_{r_1,r_2\gg r_3\gg 1}=-1,\qquad 
I_q(r_1,r_2,r_3)\big|_{r_3\gg r_1,r_2\gg 1}=-\frac{1}{2}.
\eeq
Using Eq.~\eqref{discIq} for the discontinuity, we find the 
dispersive representation
\beq
\label{disp_single}
I_q(r_1,r_2,r_3)=-1+\frac{1}{2\pi i}\int_{4}^\infty \diff x \frac{\text{disc}_{r_1} I_q(x,r_2,r_3)}{x-r_1}.
\eeq
Such a form can again be used to isolate low-energy contributions, i.e., by imposing a lower cutoff $s_u$, the limit $m_u\to 0$ can be taken, and only $I_u(r_1,r_2,r_3)=-1$ remains. For the $c$-loop contribution, the single-variable dispersion relation suggests a form
\beq
\label{Ic_single}
I_c(q_1^2,q_2^2,\mk^2)=-1+\frac{m_c^2}{\pi}\int_{s_c}^\infty\diff x \frac{2\pi\log\frac{x-q_2^2-\mk^2+\sigma_c(x)\lambda^{1/2}(x,q_2^2,\mk^2)}{x-q_2^2-\mk^2-\sigma_c(x)\lambda^{1/2}(x,q_2^2,\mk^2)}}{(x-q_1^2)\lambda^{1/2}(x,q_2^2,\mk^2)}. 
\eeq
Such a form, however, does not strictly suffice to isolate the low-energy contributions, since the second variable $q_2^2$ can still take arbitrary values. For this reason, it is advantageous to formulate a double-spectral representation and impose cutoffs in both virtualities. 

As a first step, we consider the limit $r_3=0$ of the discontinuity~\eqref{discIq}: 
\beq
\text{disc}_{r_1} I_q(r_1,r_2,0)=\frac{4\pi i}{r_1-r_2}\log\frac{1+\sigma_1}{1-\sigma_1},\qquad \sigma_1=\sqrt{1-\frac{4}{r_1}}.
\eeq
The corresponding double-spectral function becomes
\beq
\label{double_spectral_r30}
\rho(r_1,r_2,0)=2\pi^2\delta(r_1-r_2)\log\frac{1+\sigma_1}{1-\sigma_1},
\eeq
leading to 
\beq
\label{disp_double_r3_0}
I_q(r_1,r_2,0)=-1+\frac{1}{\pi^2}\int_4^\infty \diff x\int_4^\infty \diff y\frac{\rho(x,y,0)}{(x-r_1)(y-r_2)}.
\eeq
Due to the $\delta$-function in Eq.~\eqref{double_spectral_r30}, it is clear that support for the double-spectral integration only comes from the line $r_1=r_2$, which provides further motivation for imposing a cutoff only in the single-variable dispersion relation~\eqref{disp_single}.    

Next, we generalize Eq.~\eqref{double_spectral_r30} to non-zero values of $r_3$. The simplest case is $r_3<0$, admitting the representation
\begin{align}
\label{disp_double_r3_1}
 I_q(r_1,r_2,r_3)&=-1+\frac{1}{\pi^2}\int_4^\infty \diff x\int_4^\infty \diff y\frac{\rho(x,y,r_3)}{(x-r_1)(y-r_2)}\notag\\
 &=-1+\frac{1}{\pi^2}\int_4^\infty \diff x\int_{r_2^-(x,r_3)}^{r_2^+(x,r_3)}\diff y \frac{2\pi^2}{(x-r_1)(y-r_2)\lambda^{1/2}(x,y,r_3)},
\end{align}
with double-spectral function 
\beq
\rho(r_1,r_2,r_3)=\frac{2\pi^2}{\lambda^{1/2}_{123}}\theta(-\lambda_{123}-r_1r_2r_3)
\eeq
and corresponding integration boundaries
\beq
r_2^\pm(r_1,r_3)=r_1+r_3-\frac{r_1r_3}{2}\pm\frac{1}{2}\sqrt{r_1(r_1-4)r_3(r_3-4)}.
\eeq
Unfortunately, a simple representation such as Eq.~\eqref{disp_double_r3_1} only applies for $r_3<0$, otherwise, additional contributions from anomalous cuts need to be included~\cite{Lucha:2006vc,Hoferichter:2013ama,Colangelo:2015ama}. Such contributions arise because the branch points of $\lambda_{123}$, 
\beq
r_\pm(r_1,r_3) = (\sqrt{r_1}\pm \sqrt{r_3})^2,
\eeq
can move through the cut between $r_2^+$ and $r_2^-$ onto the first sheet and require a deformation of the contour. In the case $0<r_3<4$, we find the representation
\begin{align}
 \label{disp_double_r3_2}
 I_q(r_1,r_2,r_3)
 &=-1-\frac{1}{\pi^2}\int_4^\infty \diff x\frac{2\pi^2}{x-r_1}\sum_{I=\pm}\int_{r_-(x,r_3)}^{\tilde r_2^I(x,r_3)}\diff y \frac{1}{(y-r_2)\lambda^{1/2}(x,y,r_3)},
\end{align}
where 
\beq
\tilde r_2^\pm=r_1+r_3-\frac{r_1r_3}{2}\mp\frac{i}{2}\sqrt{r_1(r_1-4)r_3(4-r_3)}
\eeq
and the integration is performed along the straight line from $r_-$ to $\tilde r_2^\pm$. 
Inserting Heaviside functions $\theta(x-s_c/m_c^2)$, $\theta(|y|-s_c/m_c^2)$ into Eq.~\eqref{disp_double_r3_2}, we obtain an improved variant of Eq.~\eqref{Ic_single} that removes the low-energy contributions below $s_c$ in both virtualities. 

Finally, in the case $r_3>4$, as relevant for $q=u$, the $\tilde r_2^\pm$ become real again, but $\lambda^{1/2}(x,y,r_3)$ produces imaginary parts that correspond to the decay $K\to \bar u u$. Such imaginary parts clearly need to be covered by the low-energy part of the representation and cannot be present in the asymptotic contribution. However, the scaling of the $u$-quark integral can already be read off from Eq.~\eqref{disp_double_r3_0}: introducing a cutoff $s_u$ in both integrations and evaluating the $\delta$-function, the remaining integral becomes
\beq
I_u(q_1^2,q_2^2,0)=-1+2\int_{s_u}^\infty \frac{\diff x}{(x-q_1^2)(x-q_2^2)}m_u^2\log\frac{1+\sigma_u(x)}{1-\sigma_u(x)}
=-1+\Order(m_u^2\log m_u), 
\eeq
since $s_u$ prevents a potential infrared singularity for $q_1^2=q_2^2=0$. Given that an analysis based on Eq.~\eqref{disp_double_r3_2} yields the same scaling, we set $I_u(q_1^2,q_2^2,\mk^2)=-1$. 

\section{Integration kernels}
\label{app:kernel}

The integration kernels for the evaluation of the loop integral~\eqref{eq:RD-RA} are given as 
\begin{align}
\label{kernels}
K(x)&=\frac{2 i}{\pi^{2}\mk^2 } \int \diff^{4} k \frac{\mk^2  k^{2}-(q \cdot k)^{2}}{(q-k)^{2}\big[(p-k)^{2}-m_\ell^{2}\big]}\times\frac{x}{x-k^{2}},\\
K(x, y)&=\frac{2 i}{\pi^{2}\mk^2 } \int \diff^{4} k \frac{\mk^2  k^{2}-(q \cdot k)^{2}}{k^{2}(q-k)^{2}\big[(p-k)^{2}-m_\ell^{2}\big]}\times\frac{xy}{(x-k^{2})\big[y-(q-k)^{2}\big]},\notag\\
K(x, y,z)&=\frac{2 i}{\pi^{2} \mk^2} \int \diff^{4} k \frac{\mk^2 k^{2}-(q \cdot k)^{2}}{k^{2}(q-k)^{2}\big[(p-k)^{2}-m_\ell^{2}\big]}\times\frac{xyz}{\big(x-k^{2}\big)\big[y-(q-k)^{2}\big]\big(z-k^{2}\big)}.\notag
\end{align}
The asymptotic formula is given by~\cite{Weil:2017knt} 
\begin{align}
	\label{Aq2}
	\Re\A_\ell(\mk^2)&=\A_\ell(0)+\frac{1}{\sigma_\ell(\mk^2)}\bigg[\text{Li}_2\big[-y_\ell(\mk^2)\big]+\frac{1}{4}\log^2\big[y_\ell(\mk^2)\big]+\frac{\pi^2}{12}\bigg],\notag\\
	\A_\ell(0)&=\frac{4}{3} \int_{0}^{\infty} \diff x\bigg[(x-2) \sqrt{1+\frac{1}{x}}-x+\frac{3}{2}\bigg] \tilde c^\text{asym}(-Q^2,-Q^2),
\end{align}
where $x=Q^2/(4m_\ell^2)$. We emphasize that this approximation is only used for the asymptotic contribution, for which we verified that the incurred error is negligible compared to the uncertainty in the RG evolution. For the low-energy contributions the difference to the full evaluation of the loop integral can be sizable, and we therefore keep the exact form in terms of the kernel functions~\eqref{kernels}.

\bibliographystyle{apsrev4-1_mod_2}
\bibliography{ref}
	
\end{document}